\documentclass[11pt]{article}
\usepackage{amsmath,amsthm}
\usepackage{amssymb}
\usepackage{graphicx}
\usepackage{adjustbox}

\newcommand{\pa}{\partial}

\newcommand{\be}{\begin{equation}}
\newcommand{\ee}{\end{equation}}

\begin{document}

\title{Adiabatic soliton management: Controlling solitary wave motion while keeping the wave envelope unchanged}

\author{Robert A. Van Gorder\thanks{Department of Mathematics and Statistics, University of Otago, P.O. Box 56, Dunedin 9054, New Zealand ({rvangorder@maths.otago.ac.nz})} }
\maketitle

\begin{abstract}
We describe how to control the motion -- both speed and propagation direction -- of a nonlinear traveling wave in real time via soliton management with time-varying dispersion/diffusion and loss/gain terms. When carried out subject to certain parameter constraints we derive, the approach allows for real-time manipulation of wave motion without compromising the structure or stability of the wave, maintaining strict adiabaticity in time. This makes our approach a promising tool for physical systems which admit a degree of control through adjustable dispersion and loss/gain terms, including problems arising in nonlinear optics, atomic physics, Bose-Einstein condensates, and thermally driven chemical reactions, where one is interested in modifying the position or speed of a wave while still maintaining adiabaticity. We illustrate both the utility and simplicity of the method through several canonical examples, demonstrating how the motion of KdV solitons, Fisher-KPP solitary wavefronts, BBM solitons, nonlinear Klein-Gordon equations, and interacting nonlinear waves in a vector NLS system can be controlled in real time.
\\
\\
keywords: wavespeed control, solitons, wavefronts, dispersion management, soliton management
\end{abstract}

\section{Introduction}
The theory of traveling waves has a long and storied history, originating with the study linear waves traveling along vibrating strings \cite{d1747recherchesA, d1747recherchesB} and growing to incorporate nonlinear solitary waves \cite{russell1845report, rayleigh1876waves, boussinesq1877essai, korteweg1895xli}, with applications found in fluid mechanics and plasma dynamics, mathematical physics, optics, and theoretical biology \cite{scott2006encyclopedia}, as well as traveling wavefronts \cite{fisher1937wave, kolmogorovstudy}, with applications of these systems to genetics, ecology, physiology, chemistry, and physics \cite{grindrod1996theory}. Control of the wavespeed of a traveling wave has been shown to be useful for many applications, ranging from optics \cite{vieira2015production} to neuroscience \cite{richardson2005control, litschel2018engineering}. The most common approach is to vary the model parameters, with the constant wavespeed adjusting accordingly, although there are limitations to this approach depending on relevant parameter regimes for which waves might exist. Another option is an active control mechanism where the wavespeed is varied during transit of the wave. Despite the possible applications, the latter option is barely explored, and will be the focus of this paper. 

The ability to control the motion of a wave is of practical importance for a number of applications. Such controllability is exploited in nonlinear optics, where light has been slowed and even stopped in experiments \cite{liu2001observation, heinze2013stopped, everett2017dynamical}. Controls via dispersion management are frequently employed to modify the structure of optical waves \cite{lakoba1998conditions, ablowitz1998multiscale, biswas2001dispersion} as well as the control of atomic matter waves in Bose-Einstein condensations (BECs) \cite{eiermann2003dispersion, kevrekidis2003feshbach, kengne2021spatiotemporal}. The controlled movement of standing matter waves in a BEC has been explored experimentally \cite{gustavson2001transport, boyer2006dynamic} and recently theoretically \cite{van2021time}, with a matter wave split or moved around the domain under a time-varying external potential. Most of these control mechanisms involve changing the dispersion or loss/gain terms over time or space, and challenges related to keeping this controlled motion adiabatic were discussed in \cite{schaff2011shortcuts, van2021time}. On the macroscopic scale, reaction-diffusion systems arising in chemistry often have temperature sensitivity, both through reaction rates \cite{arrhenius1889reaktionsgeschwindigkeit, mcnaught1997compendium} and temperature-dependent diffusion \cite{hirschfelder1964molecular, mostinsky1996diffusion}. As patterns and waves occur in chemical systems under thermal forcing \cite{forbes2013thermal, van2020influence}, dispersion/soliton management via a time-varying temperature profile is yet another potential control mechanism for nonlinear waves. 

Soliton management and dispersion management are popular techniques for the modification of nonlinear waves \cite{malomed2006soliton, lakoba1998conditions, serkin2000novel, serkin2002exactly, zharnitsky2001stabilizing, sun2011soliton}, with the approach employing temporal or spatial modification of the dispersion and loss/gain terms, resulting in a non-autonomous problem \cite{serkin2007nonautonomous}. While soliton management is typically applied to problems on unbounded domains, related approaches involving a space or time varying external potential have also been used to modify the structure of matter waves on bounded domains most relevant to experiments \cite{van2020perturbation, van2021time}. Soliton management is often used to modify stability properties of a wave, making it robust against perturbations and aiding in the stable transmission of the wave. Since soliton management results in a kind of self-similar transform of the original wave variables \cite{serkin2000novel, serkin2002exactly}, the amplitude or structure of the wave envelope is often changed when the motion of the wave is modified \cite{sun2011amplification}; see, for instance, plots in \cite{porsezian2009nonautonomous}. Whether or not the wave dynamics remain adiabatic under this similarity transform will depend upon the specific management terms employed. Despite the myriad applications, soliton management was only recently used to modify the \textit{speed} - rather than just the \textit{structure} - of a traveling wave, with the method used to slow, stop, or even reverse a solitary wave in the cubic nonlinear Schr\"odinger (NLS) equation \cite{baines2018soliton}. 

The focus of this Letter is to more generally discuss the utility of wave management for controlling the motion of a traveling wave over time while preserving the shape of the wave (e.g., while maintaining adiabaticity in the strictest sense). We first outline the general method for employing soliton management to control the motion of a traveling wave, arriving at algebraic conditions on the management terms which ensure controllability. The method has three important properties:
\begin{itemize}
\item the motion (speed and direction) of the wave may be controlled via any smooth function of time;
\item the shape and structure of the wave envelope is preserved during all changes in motion (the approach maintains adiabaticity);
\item any nice properties of the equation (integrability/conditional integrability, stability) are preserved during all changes in motion.
\end{itemize}
Some of these conditions do not appear to be satisfied in other applications of soliton management, or other control strategies for waves we are aware of. In particular, standard applications of soliton management often result in a change in the wave envelope once the control acts on the wave. Other approaches, such as moving a wave with a confining potential, often result in a loss of adiabaticity. Therefore, in this Letter, we show how to obtain general conditions on the dispersion and loss/gain parameters allowing for this control; in this way, the soliton management parameters are viewed as control parameters. After describing the method, we then use the approach to control the motion of several specific families of waves: one-solitons, cnoidal wavetrains, and two-solitons under the Korteweg-de Vries (KdV) equation; vector solitons and wavetrains in a coupled NLS system; wavefronts under the Fisher-KPP equation and Lotka-Volterra system; the one-soliton solution of the Benjamin-Bona-Mahony (BBM) equation; and, a dissipative nonlinear Klein-Gordon equation. In all cases, the method of control succeeds in giving the user the ability to control the wave motion in a desired manner, while preserving the structure of the wave, thereby maintaining adiabaticity. 

\section{Wavespeed control via soliton management}\label{secgeneral}
To outline the general approach, we start with the autonomous nonlinear partial differential equation
\begin{equation}\label{mainpde}
\frac{\partial u}{\partial t} + \delta_0 L[u] + \gamma_0 N[u]=0\,,
\end{equation}
where $L$ is a differential operator (the dispersion term) on the space $\mathbb{R}^m$, and $N$ is a nonlinear differential operator (the gain or loss term), while $\delta_0$ and $\gamma_0$ are parameters which control the relative strength of each term. We allow for $N$ to contain spatial derivatives, although these should generally be lower order than the highest order derivatives in $L$. Similarly, we allow $L$ to be nonlinear, so long as it contains the highest order space derivatives. Assume that the traveling wave transform $u(\mathbf{x},t) = U(z)$, $z=\mathbf{k}\cdot \mathbf{x} - ct$, where $\mathbf{k}\in\mathbb{R}^m$ is a constant vector and $c\in\mathbb{R}$ is the wavespeed, transforms \eqref{mainpde} into the form of the ordinary differential equation 
\begin{equation}\label{odeautonomous}
-c\frac{\mathrm{d} U}{\mathrm{d}z} + \delta_0 \hat{L}[U] + \gamma_0 \hat{N}[U] =0,
\end{equation}
where the hatted operators denote analogues of $L$ and $N$ which are evaluated along the wave coordinate, $z$. If this equation admits a solution $U(z)$, then this solution is a traveling wave solution of \eqref{mainpde}. If such a solution to \eqref{odeautonomous} exists, then the wavespeed is a function of the system parameters; e.g., $c=c(\delta_0,\gamma_0)$. Viewed in terms of an inverse problem, a given wavespeed $c$ may or may not be possible, depending upon the range of $\delta_0$ and $\gamma_0$. 

Management of the autonomous equation \eqref{mainpde} requires that the parameters scaling dispersion and nonlinearity be allowed to vary in time, and the non-autonomous \textit{managed} analogue of \eqref{mainpde} reads
\begin{equation}\label{managedpde}
\frac{\partial u}{\partial t} + \delta(t) L[u] + \gamma(t) N[u]=0\,.
\end{equation}
To employ soliton management, we choose $\delta(t)$ and $\gamma(t)$ so that $u(\mathbf{x},t) = U(Z)$ where $Z = \mathbf{k}\cdot \mathbf{x} - C(t)$ for some differentiable function of time $C(t)$. The function $U$ is a solution of the relevant traveling wave ODE with $z$ replaced by $Z$, 
\begin{equation}\label{odemanaged}
-\dot{C}\frac{\mathrm{d}U}{\mathrm{d}Z} + \delta(t) \hat{L}[U] + \gamma(t) \hat{N}[U] =0, 
\end{equation}
where $\dot{C}=\frac{\mathrm{d}C}{\mathrm{d}t}$. The envelope of the wave, $U$, is the same as the solution to \eqref{odeautonomous}, and placing $u(\mathbf{x},t) = U(Z)$ into  \eqref{managedpde} results in a coupled system of parameter constraints
\begin{equation}\label{system}
\textbf{S}\left(\delta(t), \gamma(t), \dot{C}\right)=0\,,
\end{equation}
and a solution $u=U(Z)$ to \eqref{managedpde} exists provided that the system \eqref{system} admits a solution for $\delta(t)$ and $\gamma(t)$. 

We refer to \eqref{system} as the \textit{constraint system}, as it determines the solvability condition for the management parameters. Although \eqref{system} is quite general, due to the general form of the PDE specified in \eqref{managedpde}, we can still say quite a lot about the existence of an adequate collection of management parameters. The constraint system \eqref{system} depends on $\mathbf{k}$ and any other parameters present in \eqref{odemanaged}. As the functions $\delta(t)$ and $\gamma(t)$ are coefficients of the equation \eqref{odemanaged}, the constraint system \eqref{system} is always linear in $\delta(t)$ and $\gamma(t)$, hence the existence of solutions $\delta(t)$ and $\gamma(t)$ corresponding to a prescribed $C(t)$ will depend upon whether the system is full rank. This is true both for scalar equations \eqref{managedpde} and more general vector systems. The problem of finding optimal soliton management parameters for a given wavespeed profile in time therefore reduces from a complicated nonlinear PDE problem to a much simpler linear algebraic problem, provided that the original autonomous parent problem \eqref{mainpde} admits a traveling wave solution in the first place. Mathematically, the approach requires that the function $C(t)$ be at least $C^1$ in $t$. This is good enough for any realistic application: since one cannot have an infinite acceleration at any finite time, $\dot{C}$ must be bounded.

If \eqref{system} admits a solution, then the non-autonomous managed equation in \eqref{managedpde} admits the exact solution $u(x,t) = U(\mathbf{k}\cdot\mathbf{x}-C(t))$. The envelope $U$ corresponds to the solution of the autonomous equation \eqref{mainpde}, and hence the shape of the wave is unchanged under the control scheme. In this way, the control scheme maintains adiabaticity in the strictest sense. As a consequence, we can view this problem as evolving along a new  ``managed" timescale, $\tau = C(t)$. When $\delta(t)$ and $\gamma(t)$ satisfy \eqref{system}, trajectories along the timescale $\tau$ maintain integrability (or, stability) provided the parent equation \eqref{mainpde} is integrable (stable) in the original timescale $t$. In other words, for management parameters satisfying \eqref{system}, the non-autonomous managed PDE \eqref{managedpde} inherits the integrability (stability) properties of the autonomous PDE \eqref{mainpde}.

Assuming management parameters satisfying \eqref{system} exist, it is natural to explore how to choose $C(t)$ so that the motion of a given wave is controlled. A choice $C(t) = c_1 t$ for $c\neq 0$ a constant will result in a wave with constant wavespeed $c_1$ in time, as is standard. Suppose one instead desires to change the motion of a wave initially traveling with wavespeed $c_1$ so that it travels with wavespeed $c_2\neq c_1$ after time $t>T$. The seemingly obvious choice would be to consider a function of the form $C(t) = c_1$ for $t \leq T$ and $C(t)=c_2$ for $t > T$; yet, this function is not differentiable at $t=T$, resulting in a non-physical infinite acceleration. To remedy this, we must mollify the transition between distinct wavespeeds. In order to modify a wave traveling with initial wavespeed $c_1$ so that it eventually travels with wavespeed $c_2$, while keeping the function $C(t)$ smooth, we remark that any of the following functions will accomplish this goal:
\begin{subequations}\label{asymptotic}
\begin{align}
C(t) & = \frac{c_2 + c_1}{2}(t-T) + \frac{c_2 - c_1}{2}\sqrt{\left( t-T\right)^2 + X^2}\,,\\
C(t) & = \left(\frac{c_2+c_1}{2}+ \frac{c_2-c_1}{2}\tanh(X(t-T))\right)(t-T)\,,\\
C(t) & = \frac{c_2+c_1}{2}(t-T) + \frac{c_2-c_1}{2X}\log \cosh(X(t-T))\,.
\end{align}
\end{subequations}
Each of these functions scales like $C(t) \sim c_1 t$ for $t \ll T$, $C(t) \sim c_2 t$ for $t \gg T$, and remains smooth for all time. The constants $T\in \mathbb{R}$ and $X>0$ determine the time and sharpness of the change in motion. Of course, the choice of $C(t)$ is not unique, with many other examples possible. 

For all of the functions in \eqref{asymptotic}, the agreement with a fixed wavespeed of either $c_1$ or $c_2$ is asymptotic, yet in specific applications to experiments it will likely be more desirable to transition between wavespeeds over a finite time interval. Taking our motivation from bump functions which are smooth yet compactly supported, consider the transition function
\begin{equation}
\mathcal{B}(X,\tau) = \left\lbrace 1+\exp\left( \dfrac{1}{\tau} - \dfrac{1}{X-\tau} \right)\right\rbrace^{-1} \,,
\end{equation}
which is valid for $\tau \in (0,X)$ and $X>0$.
This transition function allows us to switch between two distinct behaviors smoothly, with the duration of the transition being given by the parameter $X>0$. Then, a control scheme which transitions between $n$ distinct behaviors $\mathcal{C}_1(t), \mathcal{C}_2(t), \dots , \mathcal{C}_n(t)$ in time with the $n-1$ transitions starting at $t=T_1,T_2, \dots , T_{n-1}$ is given by
\begin{equation}\label{Cgeneral}
C(t) = \begin{cases}
\mathcal{C}_1(t)\,, \quad\text{for} ~~ t \leq T_1\,,\\
\mathcal{C}_1(t) + \mathcal{B}(X_1,t-T_1)\left( \mathcal{C}_2(t) - \mathcal{C}_1(t)\right)\,, \\
\qquad\qquad\qquad\text{for} ~~ T_1 < t < T_1+X_1\,,\\
\vdots\\
\mathcal{C}_k(t)\,, \quad\text{for} ~~ T_{k-1}+X_{k-1} \leq t \leq T_k\,,\\
\mathcal{C}_k(t) + \mathcal{B}(X_k,t-T_k)\left( \mathcal{C}_{k+1}(t) - \mathcal{C}_k(t)\right)\,,\\
\qquad\qquad\qquad\text{for} ~~ T_k < t \leq T_k+X_k\,,\\
\mathcal{C}_{k+1}(t)\,, \quad \text{for} ~~ T_k+X_k \leq t \leq T_{k+1}\,,\\
\vdots\\
\mathcal{C}_{n-1}(t)\,, \quad\text{for} ~~ T_{n-2}+X_{n-2} \leq t \leq T_{n-1}\,,\\
\mathcal{C}_{n-1}(t) + \mathcal{B}(X_{n-1},t-T_{n-1})\left( \mathcal{C}_{n}(t) - \mathcal{C}_{n-1}(t)\right)\,, \\
\qquad\qquad\qquad\text{for} ~~ T_{n-1} < t \leq T_{n-1}+X_{n-1}\,,\\
\mathcal{C}_n(t)\,, \quad\text{for} ~~ t \geq T_{n-1}+X_{n-1}\,.
\end{cases}
\end{equation}
Here we have allowed for distinct transition intervals of length $X_1,X_2,\dots ,X_{n-1}>0$, although for many applications it may be simpler to take the same transition time between each distinct behavior (e.g., $X_1=X_2=\cdots =X_{n-1}=X$, for some $X>0$). As the function $C(t)$ described in \eqref{Cgeneral} is smooth, it is differentiable, and hence can be used in the constraint system \eqref{system}.

Assume one has an existing autonomous nonlinear wave one wishes to control the motion of. Our recipe for controlling the motion of a nonlinear wave can then be summarized as follows:
\begin{enumerate}
\item Extract the wave envelope from the autonomous seed equation \eqref{odeautonomous}.
\item Replace the constant wavespeed term $ct$ in the wave envelope found in Step 1 with a desired wavespeed profile $C(t)$.
\item Obtain and then solve the constraint equation \eqref{system} for the soliton management parameters in terms of $C(t)$.
\item Solve \eqref{managedpde} using the management parameters found in Step 3 to obtain the controlled wave. 
\end{enumerate}

Having outlined the general method and recipe for controlling the motion of a wave while keeping its form intact, we now move on to considering several specific examples to illustrate the general method.

\section{Control of the KdV equation}
\begin{figure}
\begin{center}
\includegraphics[width=0.4\linewidth]{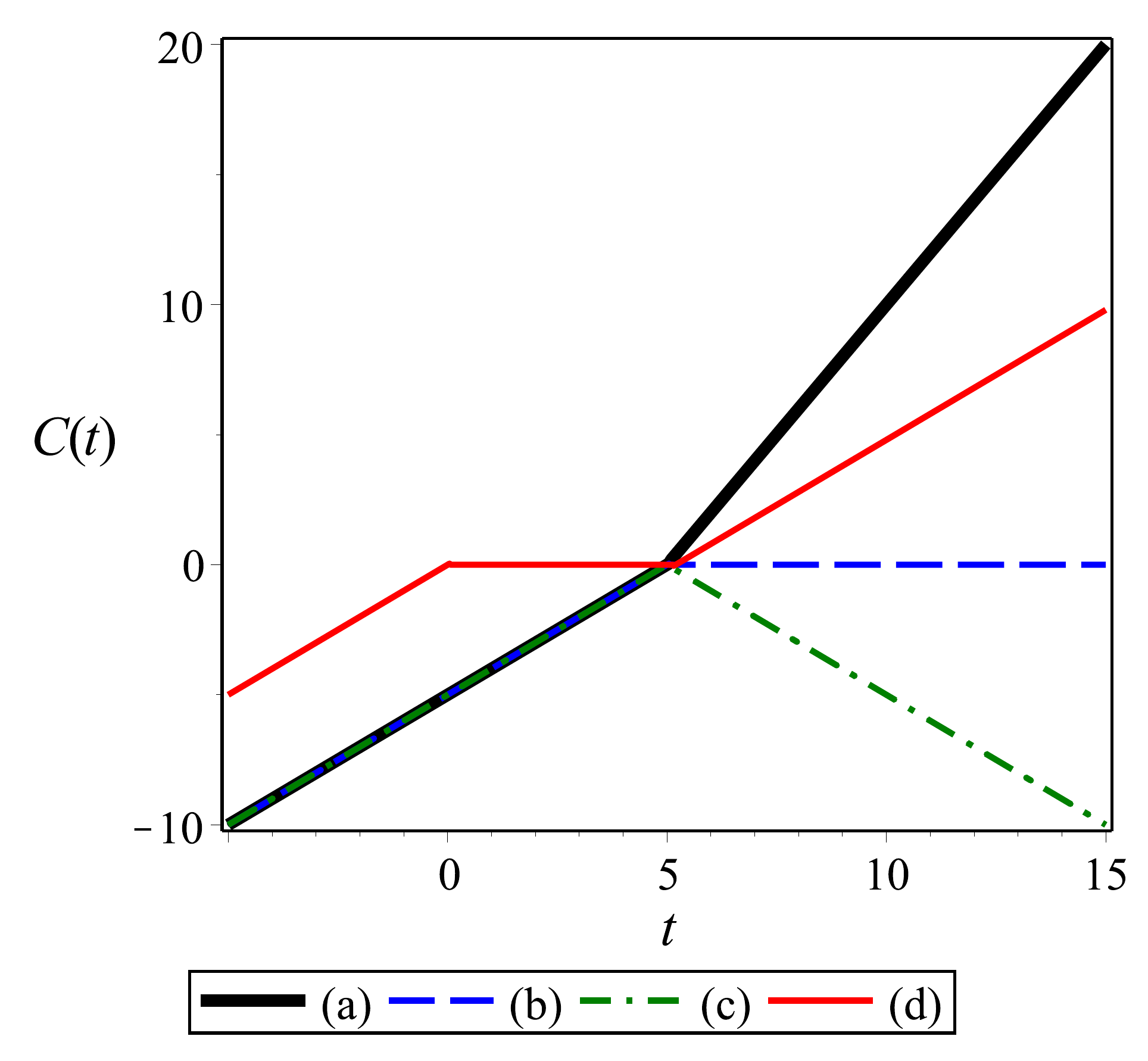}
\includegraphics[width=0.4\linewidth]{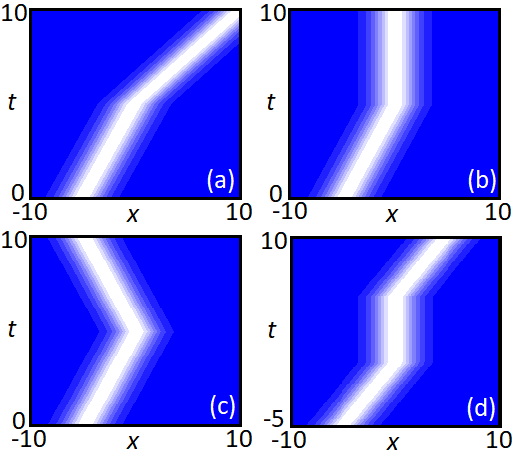}
\vspace{-0.1in}
\caption{Plots of several distinct control schemes $C(t)$ in \eqref{Cgeneral} corresponding to $n=2$, $\mathcal{C}_1(t)=c_1(t-T)$, $\mathcal{C}_2(t)=c_2(t-T)$, and $T_1=T$ giving one transition between two behaviours in (a,b,c) and $n=3$, $\mathcal{C}_1(t)=c_1(t-T_1)$, $\mathcal{C}_2(t)=0$, $\mathcal{C}_3(t)=c_2(t-T_2-2X)$ giving two transitions between three behaviours in (d). In particular, we choose (a) $c_1 =1$, $c_2=2$, (b) $c_1 =1$, $c_2=0$, (c) $c_1 = 1$, $c_2 = -1$ all with a transition at $T = 5$, and transition interval of size $X = 0.1$. These correspond to cases where the solitary wave is sped up ($c_2 >c_1$), stopped ($c_2 =0$), and reversed ($\text{sgn}(c_2)=-\text{sgn}(c_1)$) beyond time $t=T=5$. For (d), we take $c_1=c_2=1$ with $T_1 =0$, $T_2= 5$, $X=0.1$. The wave is stopped at $t=0$ for a duration of $5$ time units, before moving along with its original wavespeed. For each management choice (a-d) we plot the managed KdV solitary wave \eqref{kvdsoliton} over space interval $x\in [-10,10]$ and indicated time interval. The color scale ranges from dark blue (value of 0) to white (maximal value) on this and subsequent figures; white regions comprise the cores of respective solitary waves.\label{Fig1}}
\end{center}
\end{figure}

We first demonstrate the control technique using the Korteweg-de Vries (KdV) equation \cite{boussinesq1877essai, korteweg1895xli}, which admits the soliton solution $u(x,t)=\frac{1}{2}\text{sech}^2\left( \frac{x-t+x_0}{2}\right)$, where $x_0$ is a constant which selects the location of the peak at $t=0$ \cite{drazin1989solitons}. Consider the managed KdV equation 
\begin{equation}\label{kdvmanaged}
\frac{\partial u}{\partial t} + \delta(t)\frac{\partial^3 u}{\partial x^3} + 6\gamma(t) u\frac{\partial u}{\partial x} =0\,.
\end{equation}
Choosing $u(x,t) = U(Z)$, $Z=x-C(t)+x_0$, we have 
\begin{equation}\label{kdvode}
-\dot{C}\frac{dU}{dZ} + \delta(t)\frac{d^3U}{dZ^3} + 6\gamma(t)U\frac{dU}{dZ} = 0\,.
\end{equation}
The constraint system \eqref{system} comprises $\dot{C}=\delta(t)$, $\dot{C}=\gamma(t)$, so we choose the management parameters 
\begin{equation}\label{kdvparameters}
\delta(t) = \gamma(t) = \dot{C}(t)\,.
\end{equation}
Under \eqref{kdvparameters}, we find that \eqref{kdvmanaged} has the managed solution
\begin{equation}\label{kvdsoliton}
u(x,t) = \frac{1}{2}~\text{sech}^2\left( \frac{x-C(t) +x_0}{2}\right)\,.
\end{equation}
To illustrate the approach, we plot the exact solution \eqref{kvdsoliton} in Fig. \ref{Fig1} under four different control strategies for either slowing, stopping, or reversing the KdV soliton. As the managed KdV equation remains integrable, these solutions inherit the stability of the standard KdV soliton and hence are robust.

In addition to solitary waves and wavefronts, periodic wave trains are also controllable under the management technique. Cnoidal wave solutions, involving the Jacobi elliptic function cn, were known even to Korteweg and deVries \cite{korteweg1895xli} and have been studied in a number of subsequent works \cite{bottman2009kdv, ludu1998patterns}. Consider again the managed form of the KdV equation, and consider a solution of the form 
\begin{equation}\label{cn1}
u(x,t)=\dfrac{1}{2}\text{cn}^2\left(\dfrac{Z}{2} ,\kappa\right)\,,
\end{equation}
where the parameter $0<\kappa <1$ determines the period and structure of the cnoidal wave. Using \eqref{cn1} in \eqref{kdvode}, we find
\begin{equation}
3\left(\kappa^2\delta(t)-\gamma(t)\right)\text{sn}^2\left(\dfrac{Z}{2} ,\kappa\right)+3\gamma(t)-\left( 1+\kappa^2\right)\delta(t)-\frac{\mathrm{d}C}{\mathrm{d}t} =0 \,,
\end{equation}
hence the constraint system \eqref{system} becomes 
\begin{equation} 
\kappa^2\delta(t)-\gamma(t)=0, \quad 3\gamma(t)-\left( 1+\kappa^2\right)\delta(t)-\frac{\mathrm{d}C}{\mathrm{d}t}=0\,.
\end{equation} 
Taking the management parameters to be
\begin{equation} \label{cnparameters}
\delta(t) = \frac{1}{(2\kappa^2 -1)}\frac{\mathrm{d}C}{\mathrm{d}t}\,, \quad \gamma(t) = \frac{\kappa^2}{(2\kappa^2 -1)}\frac{\mathrm{d}C}{\mathrm{d}t}\,,
\end{equation} 
for $0 < \kappa < 1$, $\kappa \neq \pm 1/\sqrt{2}$, we find that the managed KdV equation has the exact solution 
\begin{equation} \label{cn2}
u(x,t) = \dfrac{1}{2}\text{cn}^2\left(\dfrac{x-C(t)+x_0}{2},\kappa\right)\,.
\end{equation} 
In the limit $\kappa \rightarrow 1^{-}$, the solution \eqref{cn2} is equal to the soliton solution \eqref{kvdsoliton}, and in this limit the management parameters \eqref{cnparameters} reduce to those in \eqref{kdvparameters}. Note that the case where $\kappa = \pm 1/\sqrt{2}$ gives only a standing wavetrain, even for the autonomous problem. As such, the wavetrain is not controllable since only a stationary wave rather than a traveling wave exists for the autonomous problem. This is therefore not a direct failing of the management technique, but simply a result of the non-existence of a traveling wave solution to the autonomous problem.

\begin{figure}
\begin{center}
\begin{tabular}{ccc} 
Time &Standard KdV Two-Soliton & Managed KdV Two-Soliton \\
\adjustbox{valign=c}{$t=-0.5$}&\adjustbox{valign=c}{\includegraphics[width=0.4\textwidth]{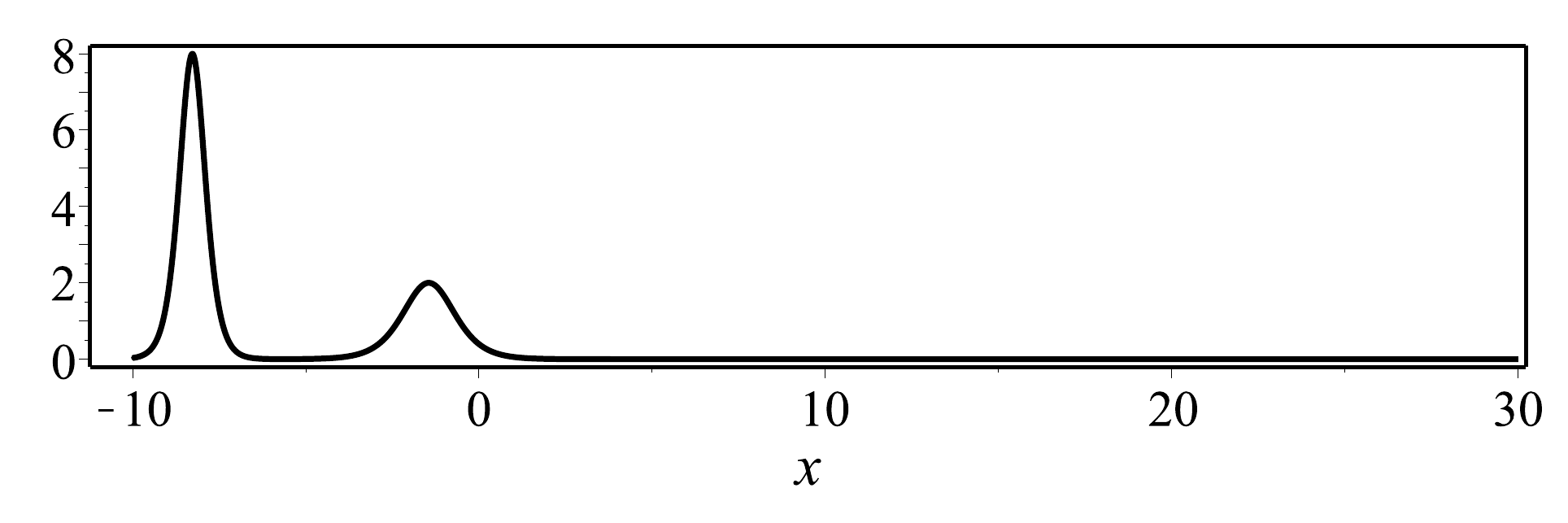}} & \adjustbox{valign=c}{\includegraphics[width=0.4\textwidth]{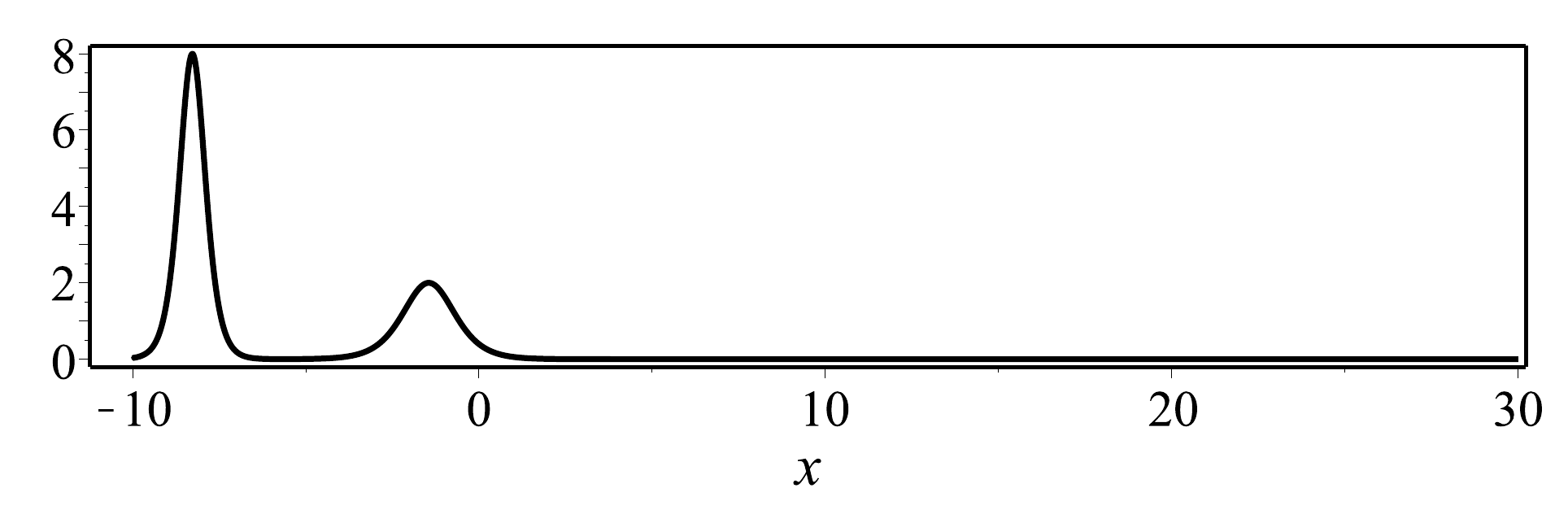}}\\
\adjustbox{valign=c}{$t=-0.2$}&\adjustbox{valign=c}{\includegraphics[width=0.4\textwidth]{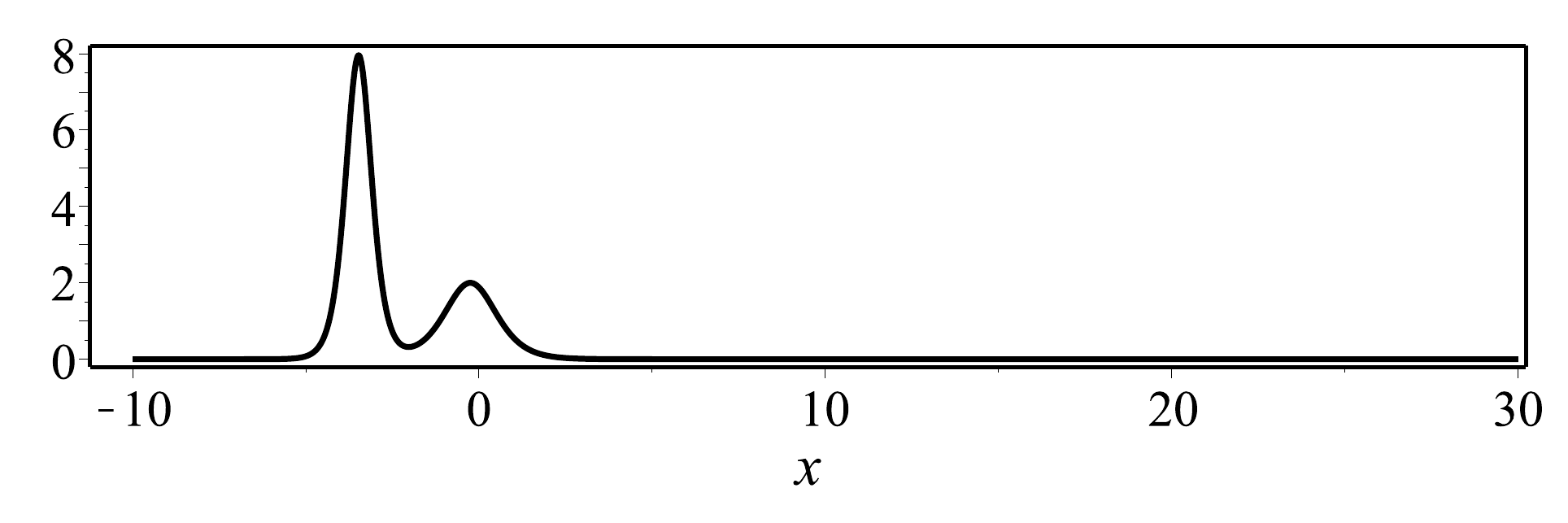}} & \adjustbox{valign=c}{\includegraphics[width=0.4\textwidth]{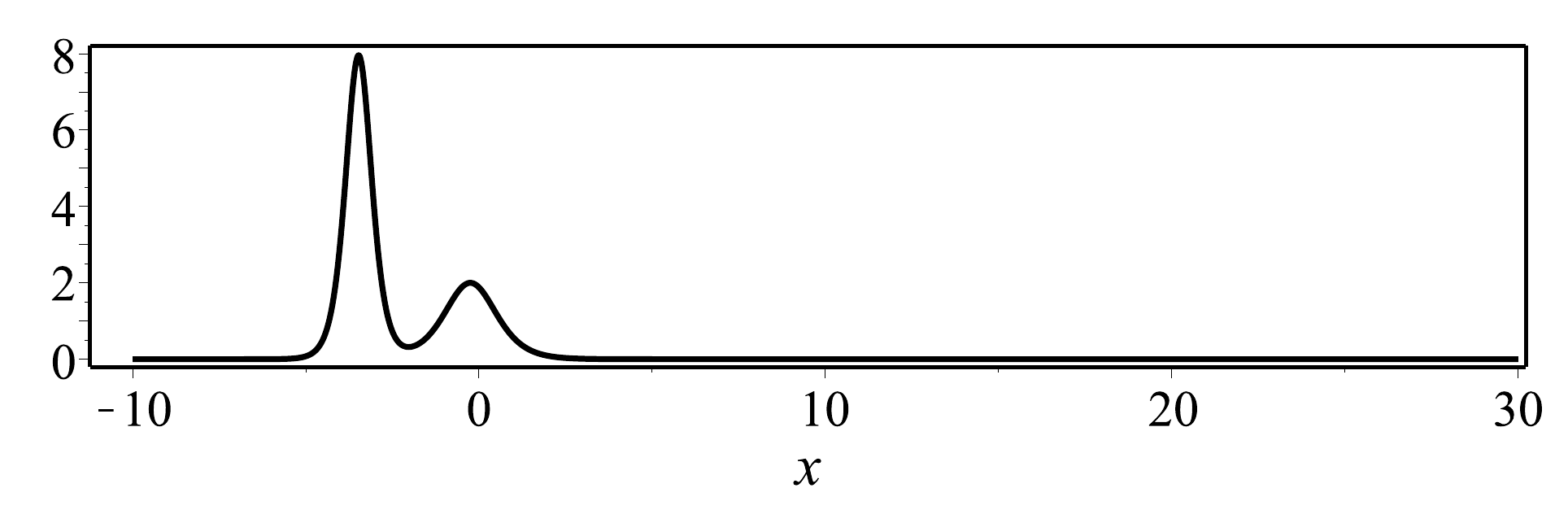}}\\
\adjustbox{valign=c}{$t=0.0$}&\adjustbox{valign=c}{\includegraphics[width=0.4\textwidth]{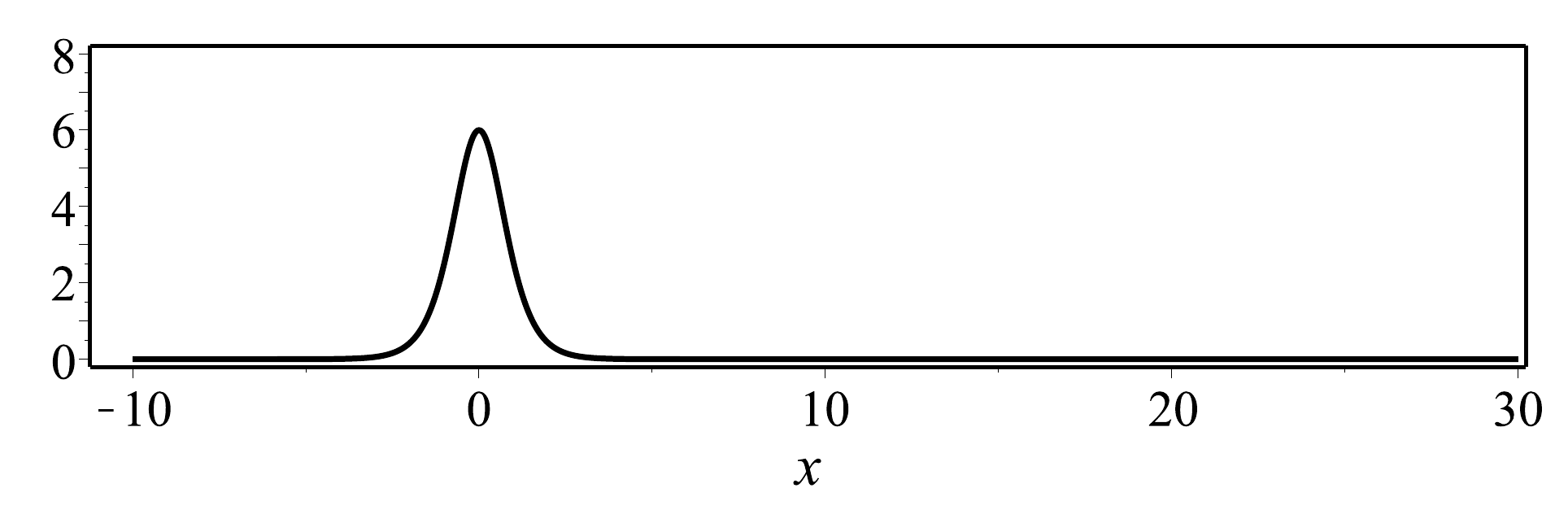}} & \adjustbox{valign=c}{\includegraphics[width=0.4\textwidth]{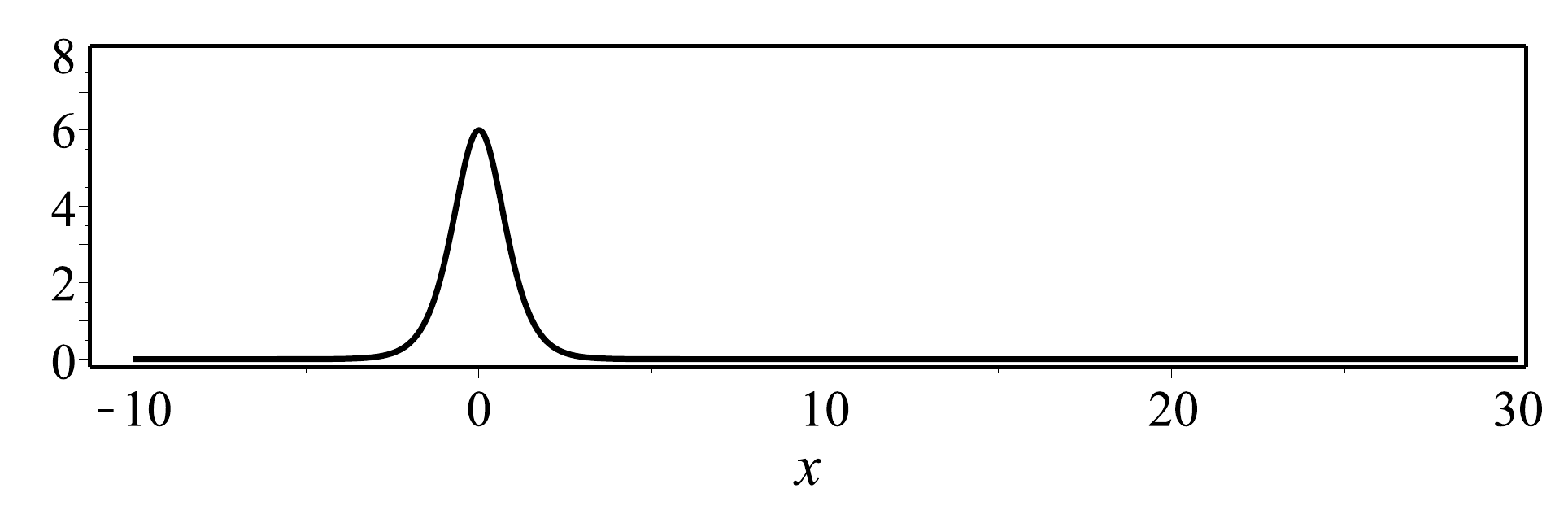}}\\
\adjustbox{valign=c}{$t=0.2$}&\adjustbox{valign=c}{\includegraphics[width=0.4\textwidth]{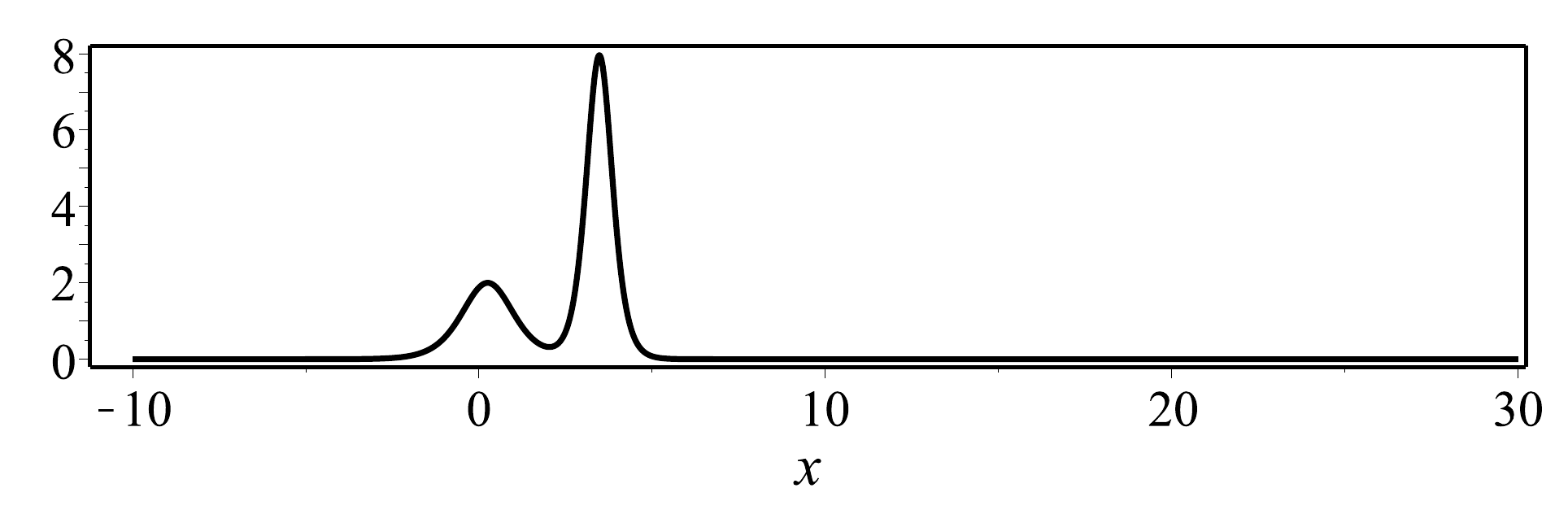}} & \adjustbox{valign=c}{\includegraphics[width=0.4\textwidth]{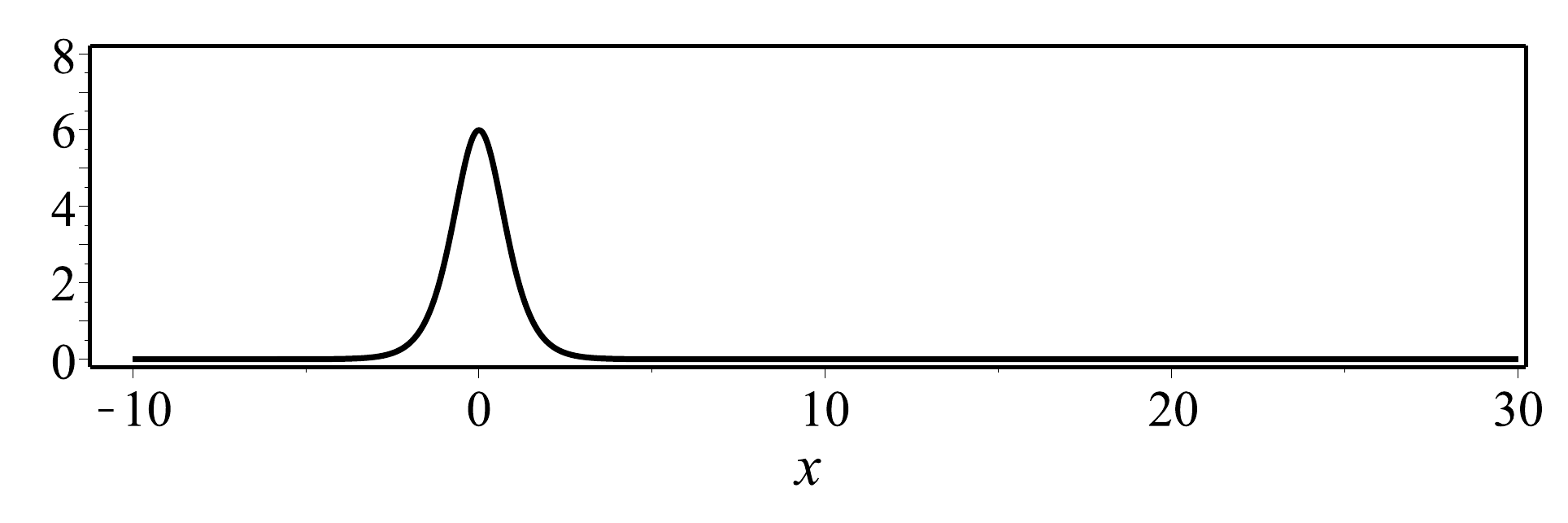}}\\
\adjustbox{valign=c}{$t=0.5$}&\adjustbox{valign=c}{\includegraphics[width=0.4\textwidth]{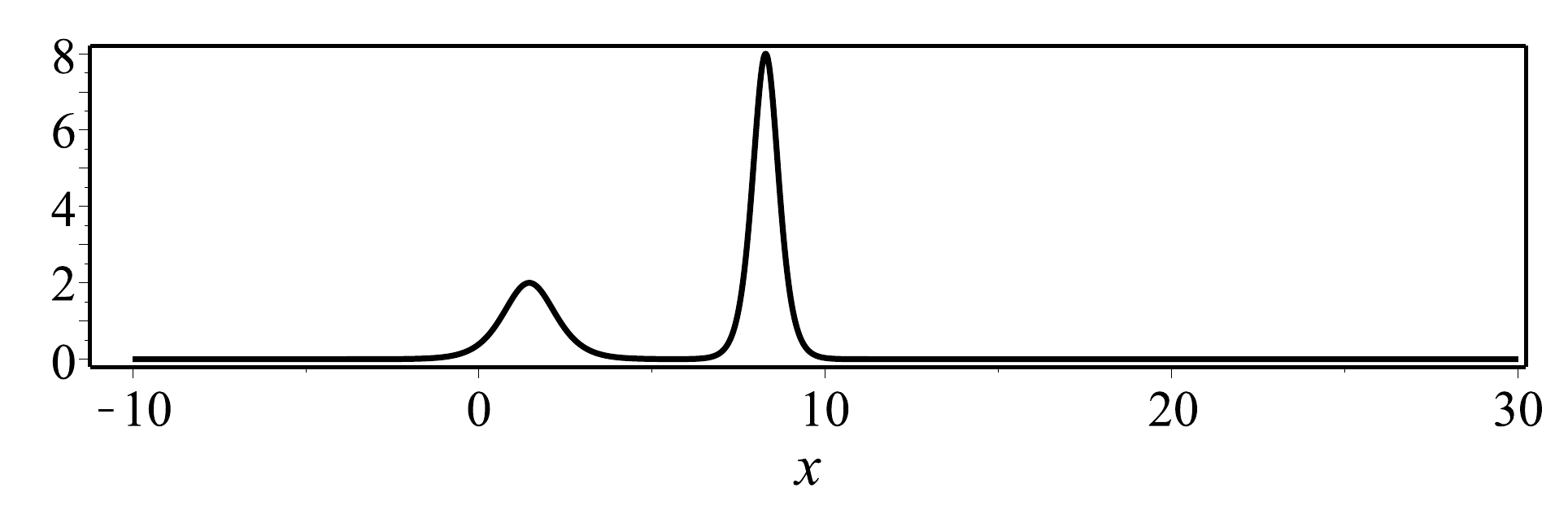}} & \adjustbox{valign=c}{\includegraphics[width=0.4\textwidth]{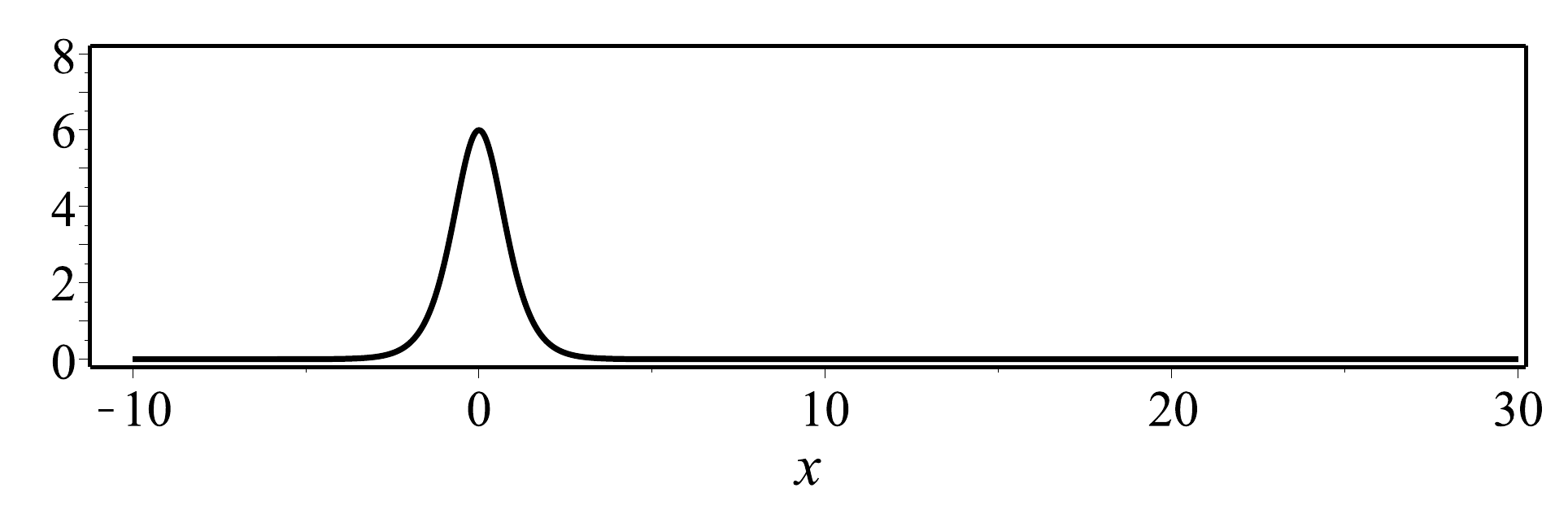}}\\
\adjustbox{valign=c}{$t=1.0$}&\adjustbox{valign=c}{\includegraphics[width=0.4\textwidth]{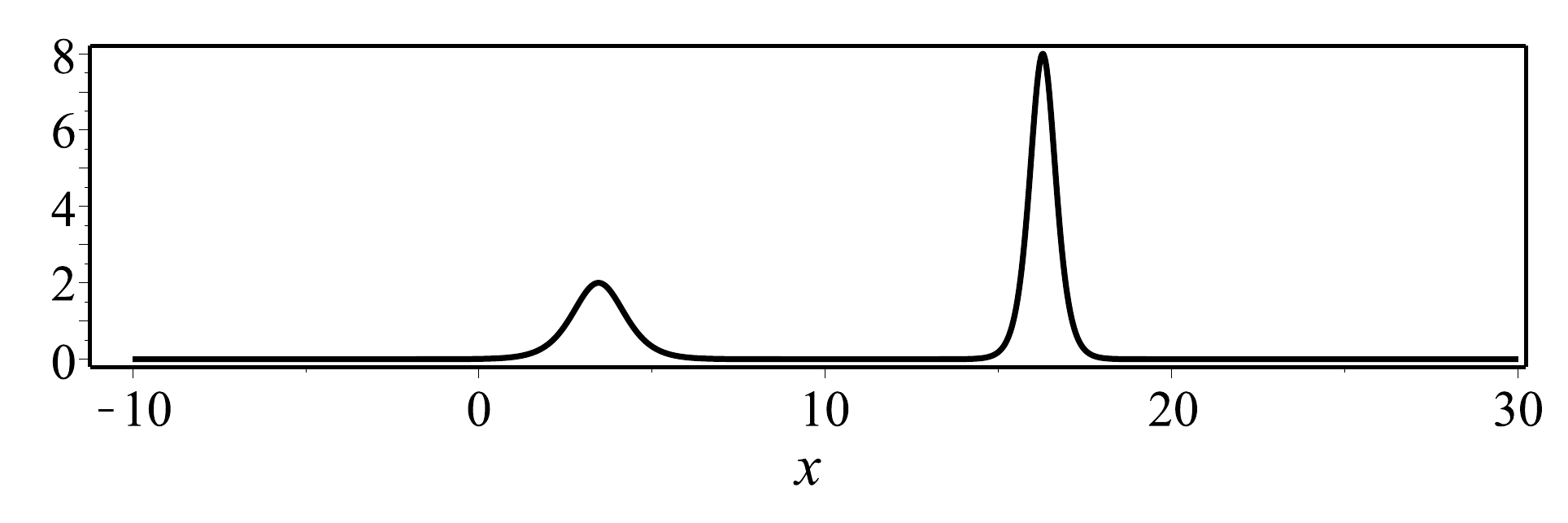}} & \adjustbox{valign=c}{\includegraphics[width=0.4\textwidth]{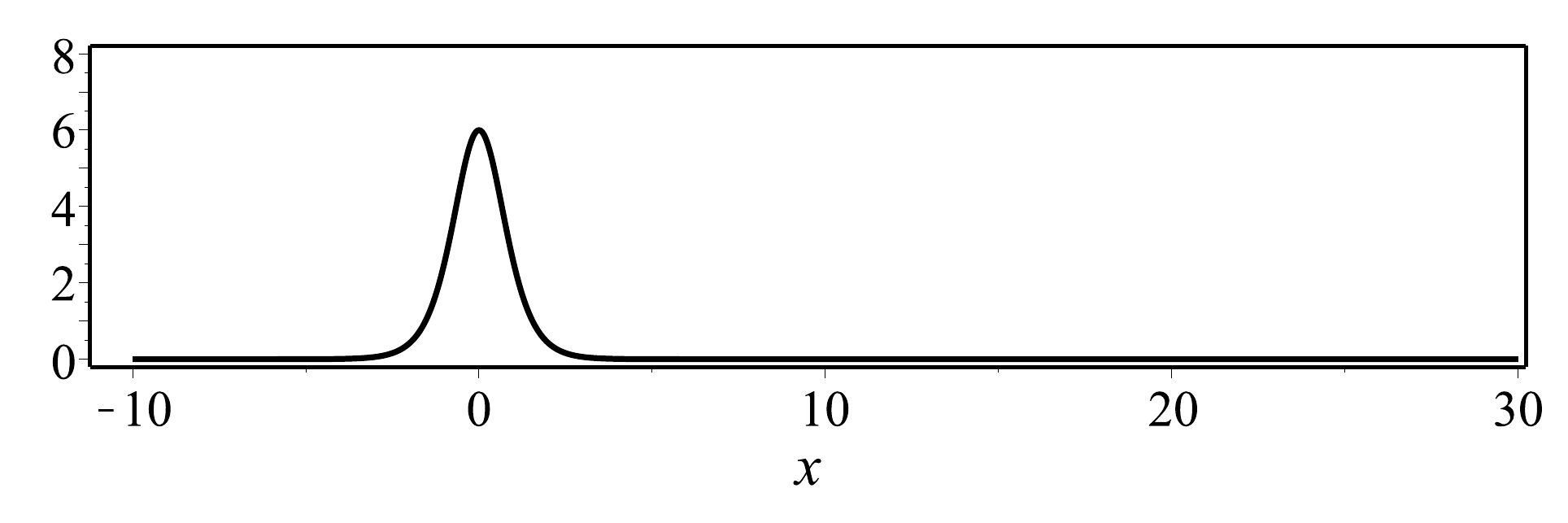}}\\
\adjustbox{valign=c}{$t=1.5$}&\adjustbox{valign=c}{\includegraphics[width=0.4\textwidth]{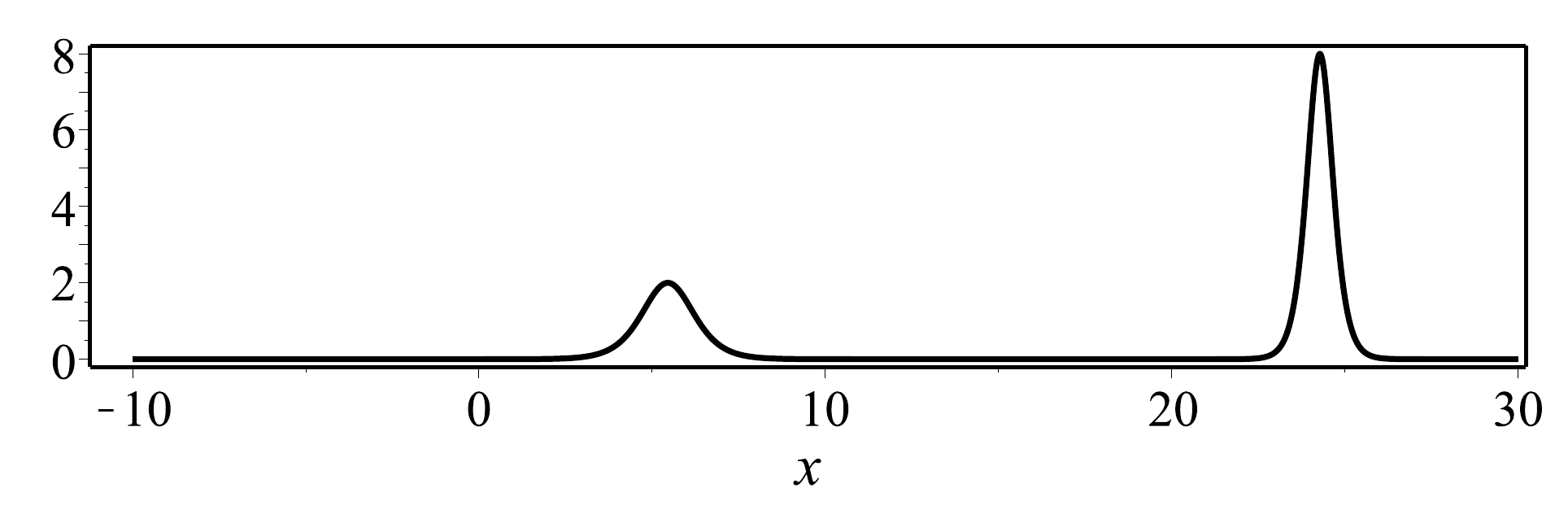}} & \adjustbox{valign=c}{\includegraphics[width=0.4\textwidth]{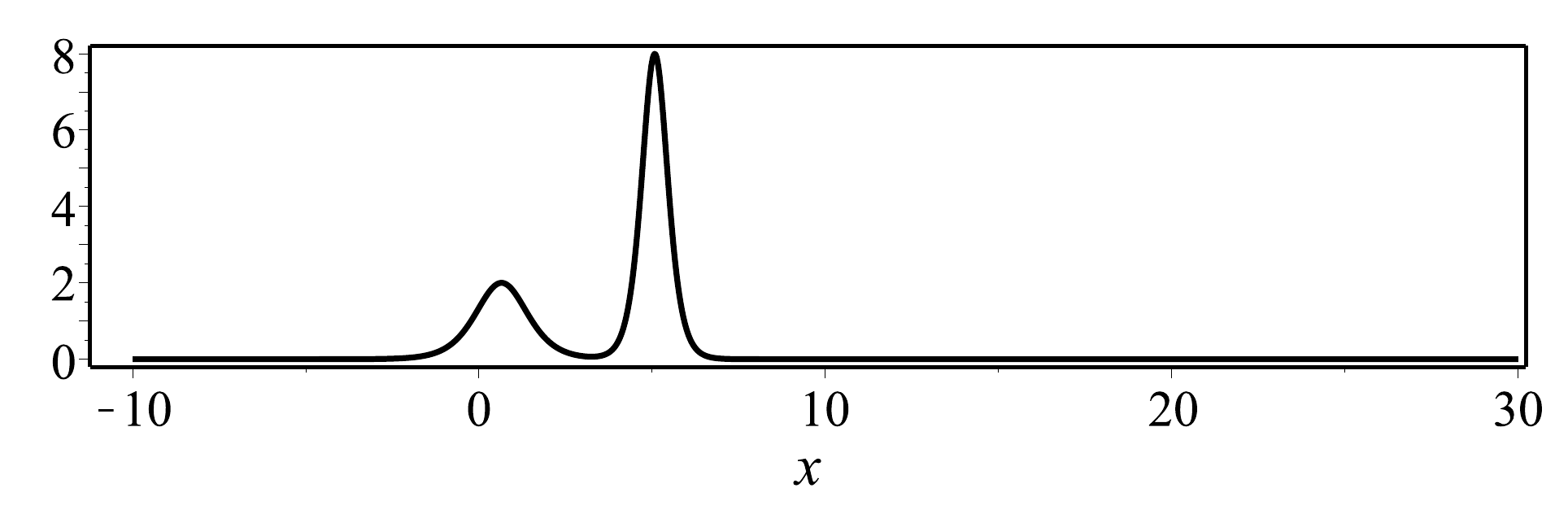}}\\
\adjustbox{valign=c}{$t=2.0$}&\adjustbox{valign=c}{\includegraphics[width=0.4\textwidth]{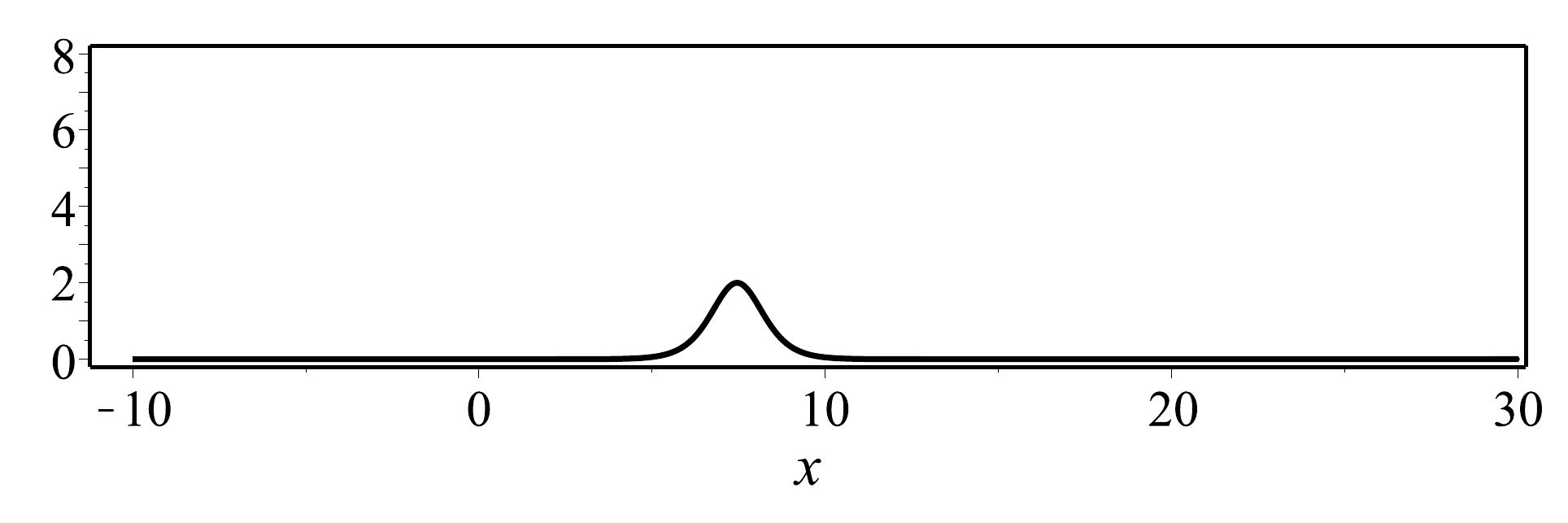}} & \adjustbox{valign=c}{\includegraphics[width=0.4\textwidth]{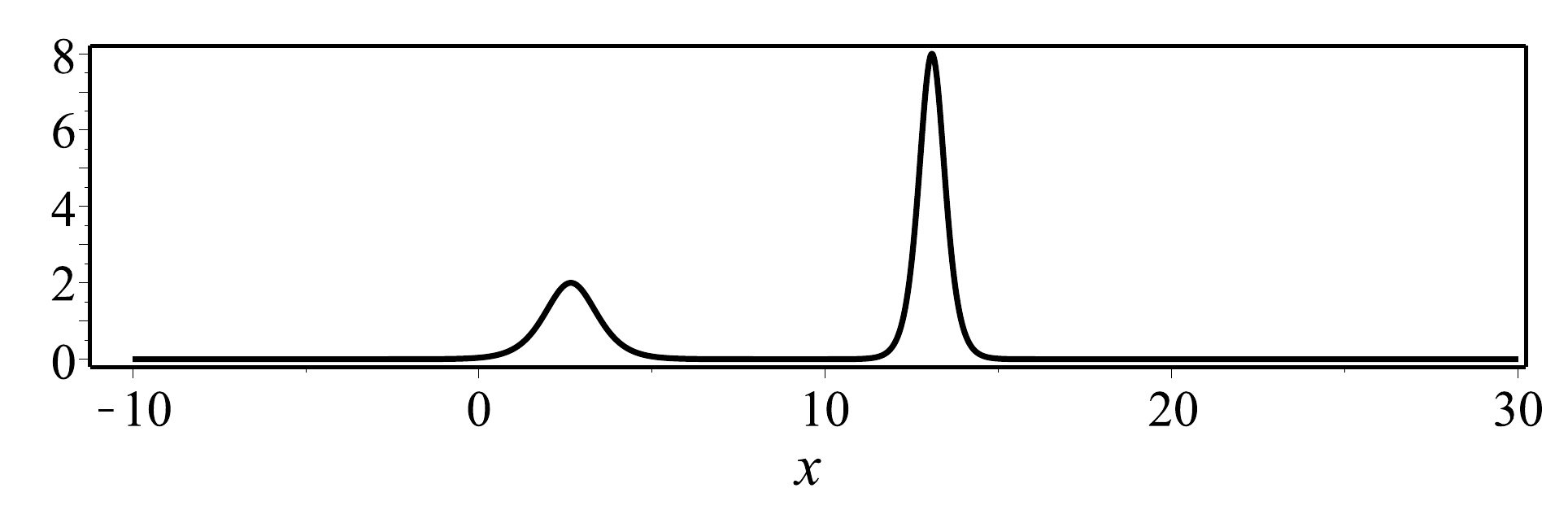}}\\
\end{tabular}
\vspace{-0.15in}
\caption{Plot of the standard (left) and managed (right) two-soliton solution of the KdV equation. We choose a management technique \eqref{Cgeneral} which freezes the motion of the solitons for one time unit starting at $t=0$, taking $c_1=c_2=1$, $T_1=0$, $T_2=1$, and $X=0.1$. We plot time snapshots of the solutions over space: from top to bottom, we plot each of the solutions at times $t=-0.5, -0.2,0,0.2,0.5,1.0,1.5,2.0$. The result of the management is that the managed soliton retains the one-hump configuration for one time unit before the two solitons separate. After separation, the solitons proceed with their separate motion (as standard), and are undisturbed due to the management. \label{FigInteract}}
\end{center}
\end{figure}

In addition to the one-soliton and cnoidal wavetrains, it is also possible to study soliton interactions using our method. Suppose one wishes to slow the interaction of two or more solitons. To model soliton-soliton interactions under the KdV equation, we consider the two-soliton solution \cite{zabusky1965interaction, gardner1967method}. Going through the standard derivations using the managed form of the KdV equation \eqref{kdvmanaged}, we find that the managed two-soliton solution
\begin{equation}
u(x,t) = 12\dfrac{3+4\cosh\left(2[x-4C(t)]+\cosh\left(4[x-16C(t)]\right)\right)}{\left( 3\cosh\left(x-28C(t)\right) + \cosh^3\left(3[x-9C(t)]\right) \right)^2}
\end{equation}
exists provided the management parameters again satisfy \eqref{kdvparameters}. We plot one example of a managed two-soliton interaction in Fig. \ref{FigInteract}. We show that management can be used to prolong the interaction of the solitons, keeping both solitons together when they would normally separate after collision. After one time unit, the solitons are finally allowed to go on their way, and they move the same as standard un-managed solitons after an interaction, as if they were never modified in the first place.

\begin{figure}
\begin{center}
\includegraphics[width=0.5\textwidth]{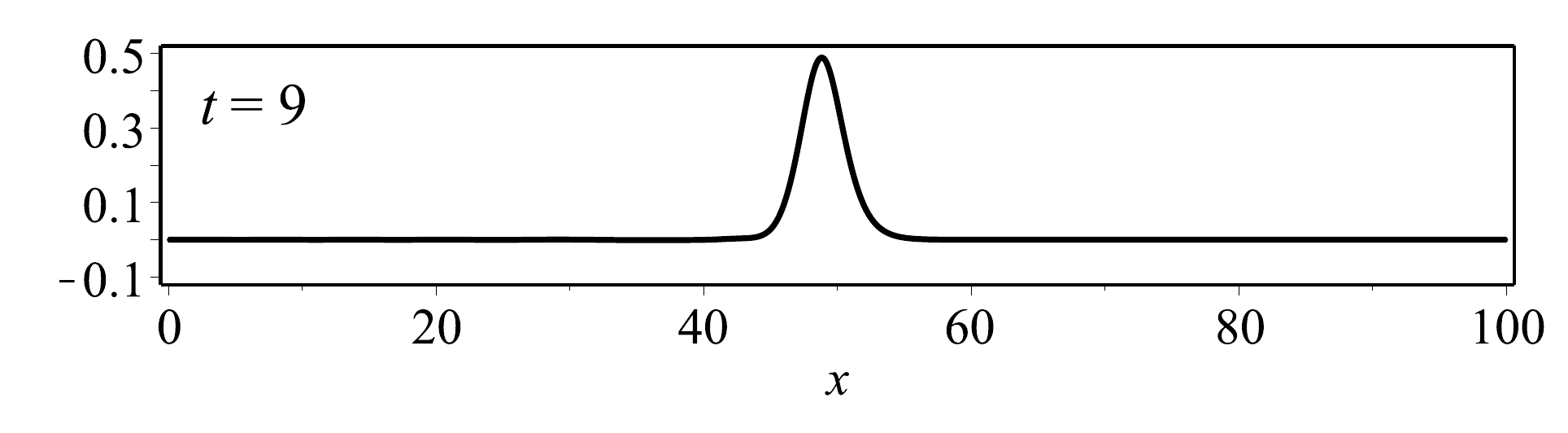}\\
\includegraphics[width=0.5\textwidth]{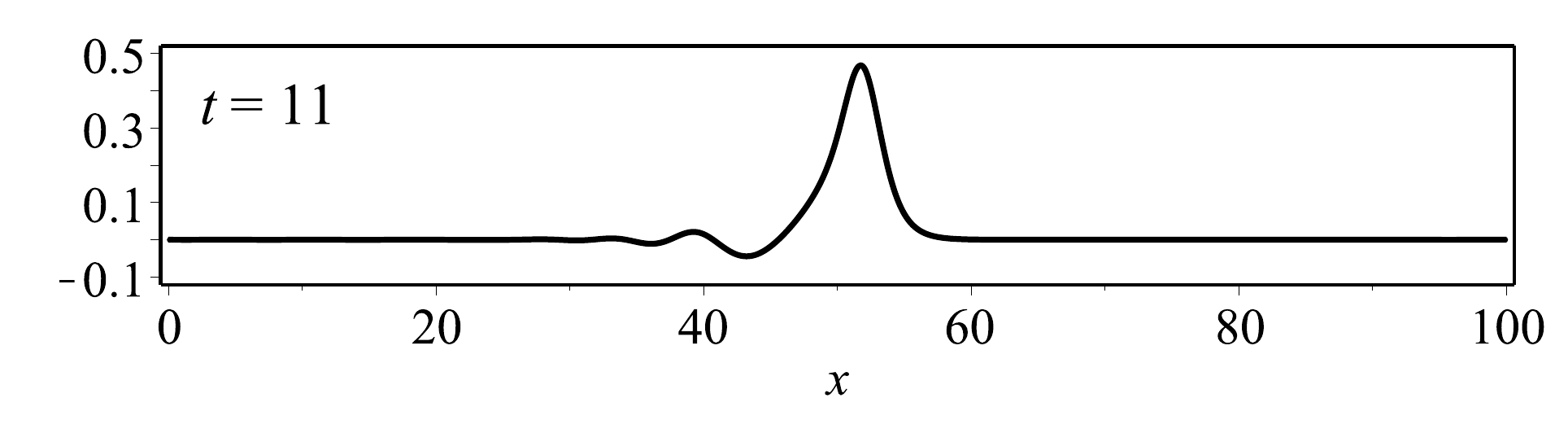}\\
\includegraphics[width=0.5\textwidth]{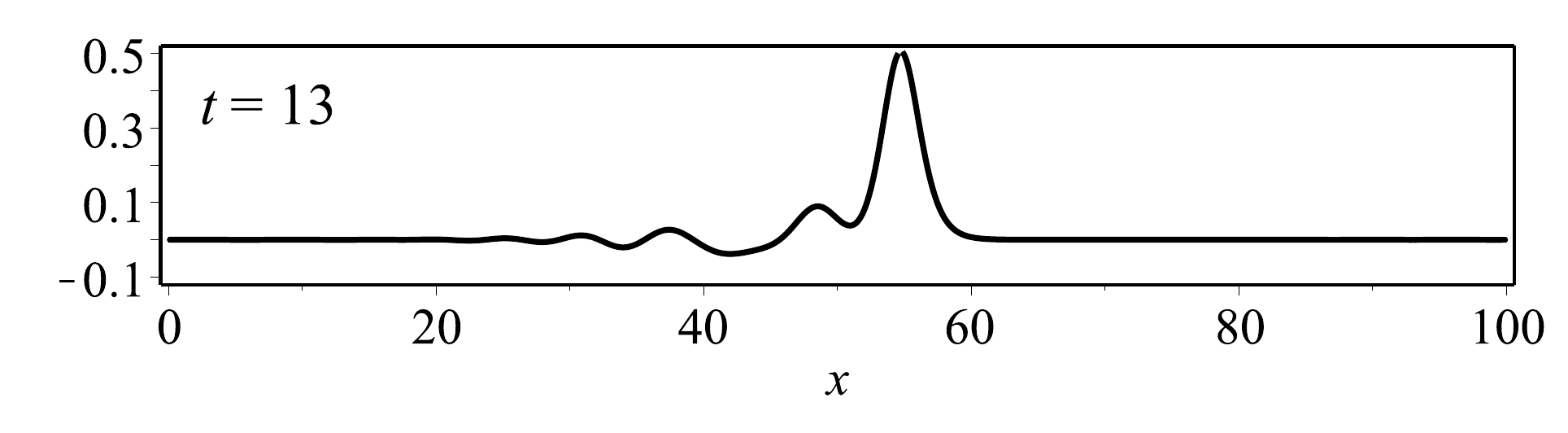}\\
\includegraphics[width=0.5\textwidth]{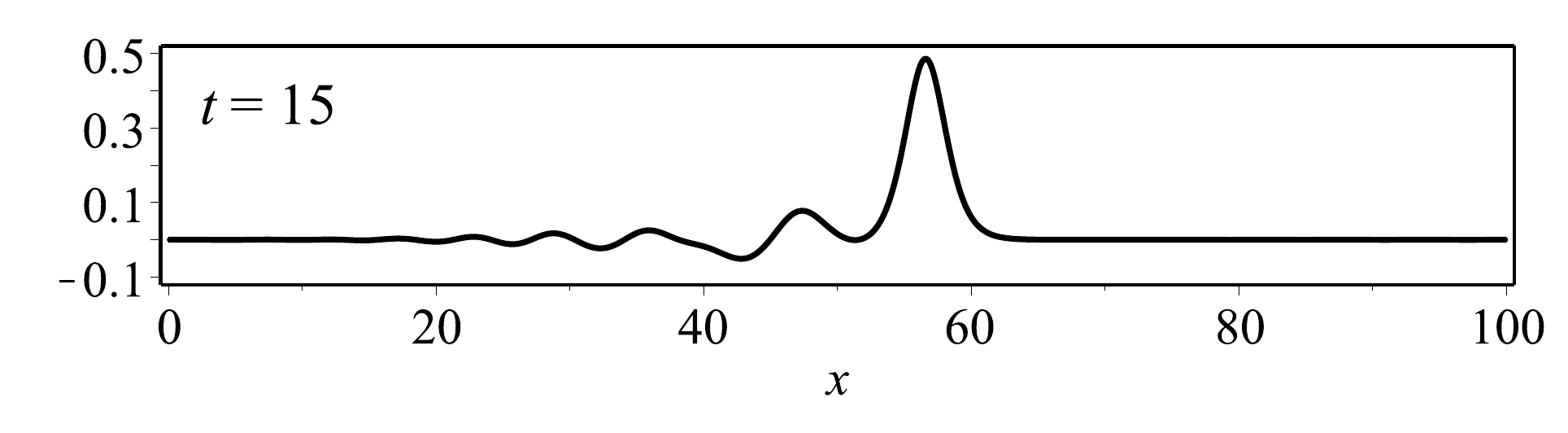}\\
\includegraphics[width=0.5\textwidth]{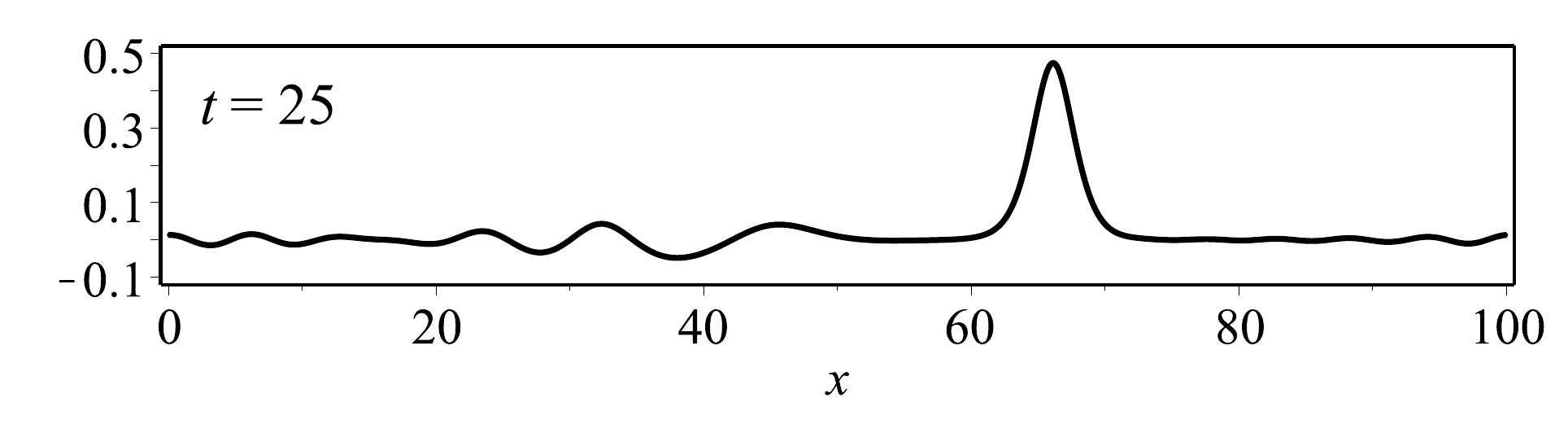}
\vspace{-0.2in}
\caption{Failure of KdV soliton to remain adiabatic in light of ``bad" management parameters not satisfying \eqref{system}. The choice of soliton management parameters which do not satisfy the constraint system \eqref{system} results in failure of the approach, leading to non-adiabatic evolution which distorts the wave envelope leading to radiation of mass from the primary wave. We choose the managed KdV equation \eqref{kdvmanaged} with management parameters $\delta(t)=1+\exp(-(t-10)^2)$ and $\gamma(t) = 1+\exp(-(t-11)^2)$, which are incompatible with \eqref{kdvparameters}. Although the management parameters are out of sync for only a brief time, the damage is done, and the shed mass radiates away from the core of the soliton. \label{FigBadKdV}}
\end{center}
\end{figure}

Having shown the utility of our approach to soliton management for controlling various solutions of the KdV equation, it is of course natural to wonder if our conditions \eqref{kdvparameters} on the management parameters can be relaxed. We show an example of what happens when management parameters do not obey the system \eqref{system} in Fig. \ref{FigBadKdV}, where the dispersion and loss/gain parameters are out of sync by one time unit. The solution is obtained numerically via the finite element method software FLEX PDE \cite{flexpde2017solutions}. The temporary mismatch in the parameters results in the gradual destruction of the one-soliton, and the soliton continues to collapse and radiate mass even after the two parameters come into alignment again; by this point, the damage has been done. This example demonstrates why it is essential to use our approach outlined in Sec. \ref{secgeneral}. Although the approach we outline is quite simple, it is powerful, and ad hoc selection of management parameters ignoring our approach will most likely end in failure of adiabaticity.

\section{Control of vector wave systems}
\begin{figure}
\begin{center}
\includegraphics[width=0.4\linewidth]{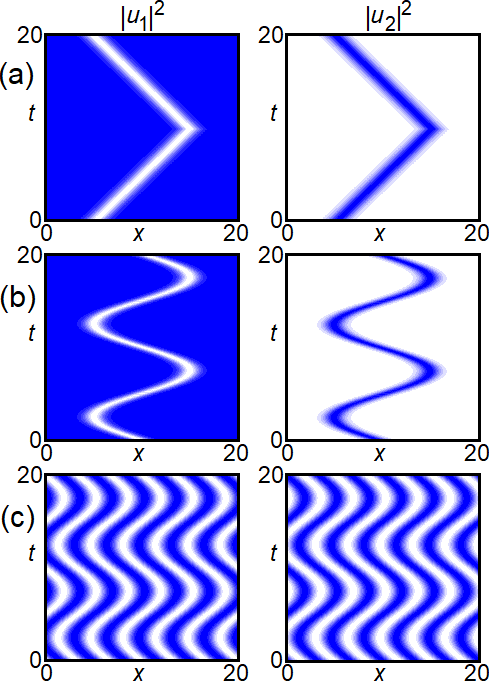}
\vspace{-0.15in}
\caption{(a,b) Managed bright-dark vector soliton solutions \eqref{vectnlssoln} to the NLS system \eqref{coupledNLS}. We plot the square modulus of $u_1$ and $u_2$, assuming (a) $C(t)$ takes the form \eqref{Cgeneral} with $n=2$, $\mathcal{C}_1(t)=t-T_1$, $\mathcal{C}_2=-(t-T_1)$, $T_1=10$, $X_1=0.01$, $x_0=-15$ and (b) $C(t)=-5\sin(\pi t/5)$, $x_0=-10$. (c) Managed cnoidal-snoidal wavetrain solution \eqref{vectwavetrains} using $C(t)=-2\sin(\pi t/5)$, $x_0=-10$. Colors are as indicated in Fig. \ref{Fig1}. \label{Fig3}}
\end{center}
\end{figure}

It is a straightforward matter to extend our approach for controlling wave motion to systems of two or more equations. In this setting, the quantities $u$, $L$, $N$ in \eqref{mainpde} and \eqref{managedpde} are viewed as vectors, with $\delta$ and $\gamma$ viewed as diagonal matrices. The entire procedure will be the same, with the constraint system \eqref{system} involving all entries of $\delta$ and $\gamma$, along with derivatives of $C$. As an example, consider the managed coupled NLS system
\begin{equation}\label{coupledNLS}
\begin{aligned}
& \mathrm{i}\frac{\pa u_1}{\pa t} + \delta_1(t) \frac{\pa^2 u_1}{\pa x^2} + \gamma_1(t) \left( |u_1|^2 + \alpha |u_2|^2\right)u_1 =0\,,\\
& \mathrm{i}\frac{\pa u_2}{\pa t} + \delta_2(t) \frac{\pa^2 u_2}{\pa x^2} + \gamma_2(t) \left( \beta |u_1|^2 + |u_2|^2\right)u_2 =0\,,
\end{aligned}\end{equation}
where $0< \alpha, \beta < 1$ are cross-phase modulation parameters. Systems of this form are known to permit vector solitons in the autonomous case, with bright-bright, dark-dark, or bright-dark pairs possible \cite{menyuk1987nonlinear, kivshar1993vector, yang2000fractal}. Considering bright-dark vector solutions of the form $u_1 = \exp(\mathrm{i}[x-\Omega_1(t)])\text{sech}(Z)$, $u_2 = \exp(\mathrm{i}[x-\Omega_2(t)])\tanh(Z)$, we obtain the constraint system $\dot{\Omega}_1 + \gamma_1=2\delta_1$, $\dot{\Omega}_2 + \beta \gamma_2 = 3\delta_2$, $\dot{C}= 2\delta_1$, $\dot{C} = 2\delta_2$, $(1-\alpha) \gamma_1 = 2\delta_1$, $(1-\beta)\gamma_2 =- 2\delta_2$. Solving this system of six equations assuming $\alpha \neq 1$ and $\beta \neq 1$, we obtain the management parameters
\be
\delta_1=\delta_2 = \frac{1}{2}\dot{C},~ \gamma_1 = \frac{1}{1-\alpha}\dot{C},~  \gamma_2 = -\frac{1}{1-\beta}\dot{C},
\ee
as well as the time-varying spectral parameters
\be 
\Omega_1 = -\frac{\alpha}{1-\alpha}C, \quad \Omega_2= \frac{3-\beta}{2(1-\beta)}C,
\ee
resulting in the managed bright-dark soliton pair
\begin{align}
u_1(x,t)&=\exp\left(\mathrm{i}\left[x+\frac{\alpha C(t)}{1-\alpha}\right]\right)\text{sech}(x-C(t)+x_0)\,,\nonumber\\
u_2(x,t)&=\exp\left(\mathrm{i}\left[x-\frac{(3-\beta)C(t)}{2(1-\beta)}\right]\right)\tanh(x-C(t)+x_0)\,.\label{vectnlssoln}
\end{align}
We plot solutions \eqref{vectnlssoln} under two types of wavespeed controls -- one with the motion of the waves reversed, and one with a periodic back-and-forth motion -- in Fig. \ref{Fig3}(a,b).

In addition to solitary waves and wavefronts, periodic wavetrains are also controllable under soliton management. Cnoidal waves exist for the NLS equation \cite{bronski2001bose, mallory2013stationary, mallory2014stationary}, motivating us to consider managed vector wavetrain solutions to \eqref{coupledNLS}. We consider cnoidal-snoidal vector solutions of the form $u_1 = \exp(\mathrm{i}[x-\Omega_1(t)])\text{cn}(Z,\kappa)$, $u_2 = \exp(\mathrm{i}[x-\Omega_2(t)])\text{sn}(Z,\kappa)$, where cn and sn denote the respective Jacobi sine and cosine elliptic functions, while $0\leq \kappa^2 \leq 1$ denotes the elliptic modulus. We obtain the constraint system $\dot{\Omega}_1 + \gamma_1=2\delta_1$, $\dot{\Omega}_2 + \beta \gamma_2 = (2+\kappa^2)\delta_2$, $\dot{C}= 2\delta_1$, $\dot{C} = 2\delta_2$, $(1-\alpha) \gamma_1 = 2\kappa^2\delta_1$, $(1-\beta)\gamma_2 =- 2\kappa^2\delta_2$, and upon solving this system we find the management and spectral parameters
\begin{align}
&\delta_1=\delta_2 = \frac{1}{2}\dot{C},~ \gamma_1 = \frac{\kappa^2}{1-\alpha}\dot{C},~  \gamma_2 = -\frac{\kappa^2}{1-\beta}\dot{C},\nonumber\\ 
&\Omega_1 = -\frac{\alpha-(1-\kappa^2)}{1-\alpha}C, ~ \Omega_2= \frac{2+\kappa -(2-\kappa)\beta}{2(1-\beta)}C.
\end{align}
The managed vector wavetrains then take the form
\begin{align}
u_1(x,t)&=\exp\left\lbrace\mathrm{i}\left[x+\frac{\alpha-(1-\kappa^2)}{1-\alpha}C\right]\right\rbrace\text{cn}(x-C+x_0,\kappa)\,,\nonumber\\
u_2(x,t)&=\exp\left\lbrace\mathrm{i}\left[x-\frac{2+\kappa -(2-\kappa)\beta}{2(1-\beta)}C\right]\right\rbrace\text{sn}(x-C+x_0,\kappa)\,.\label{vectwavetrains}
\end{align}
When $\kappa=1$, the solutions \eqref{vectwavetrains} reduce to \eqref{vectnlssoln}. When $\alpha = \beta =0$, the solutions \eqref{vectwavetrains} reduce to the respective cnoidal and snoidal solutions of the scalar NLS equation. We plot the pair of solutions \eqref{vectwavetrains} in Fig. \ref{Fig3}(c), choosing a management scheme which moves the wavetrains back and forth in a periodic manner. 

\section{Control of reaction-diffusion wavefronts}
\begin{figure}
\begin{center}
\includegraphics[width=0.3\textwidth]{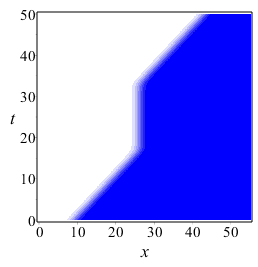}
\includegraphics[width=0.6\textwidth]{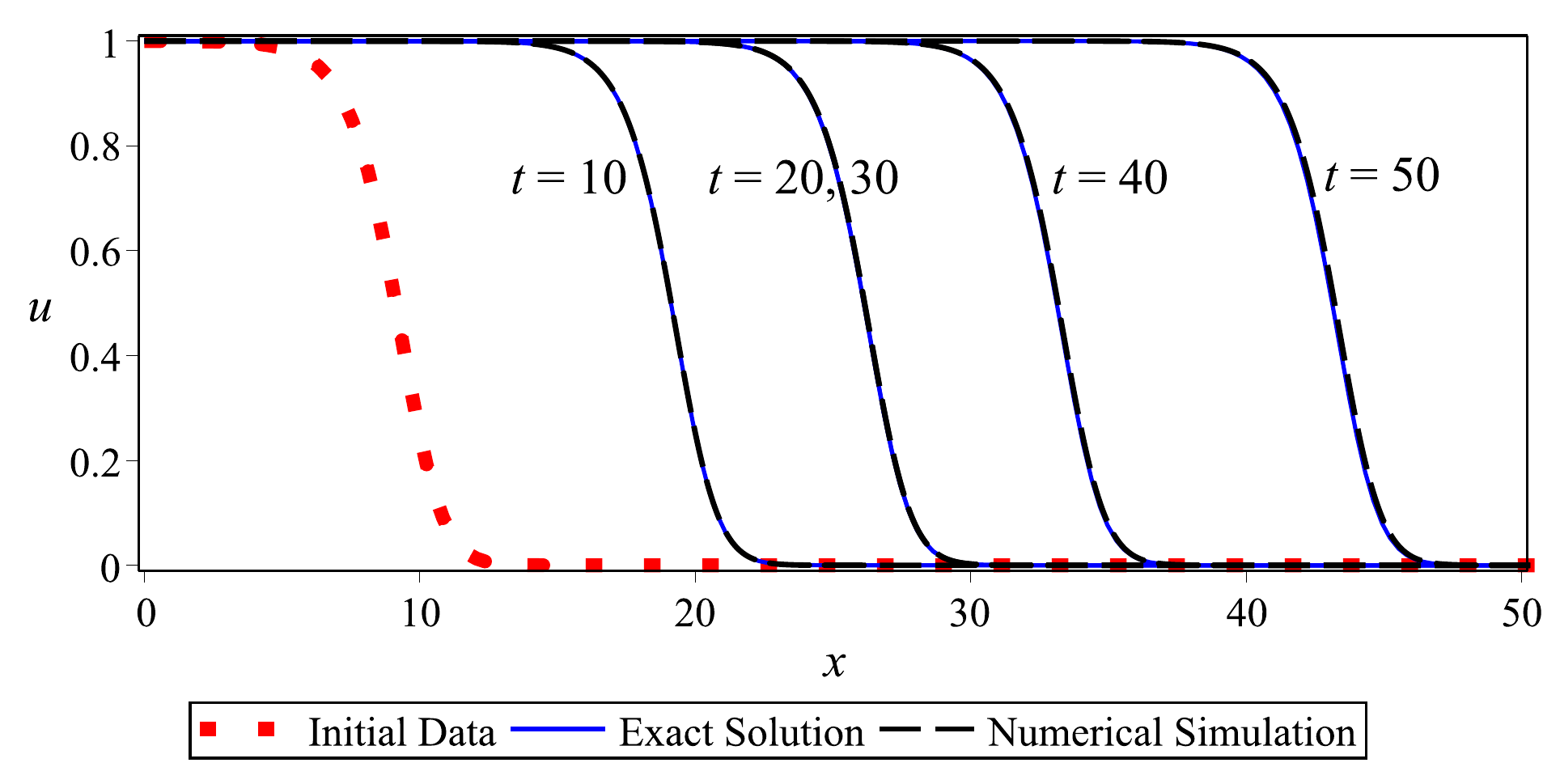}
\includegraphics[width=0.3\textwidth]{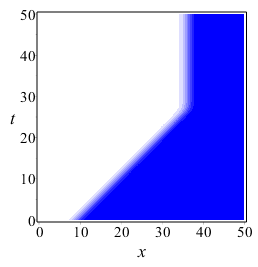}
\includegraphics[width=0.6\textwidth]{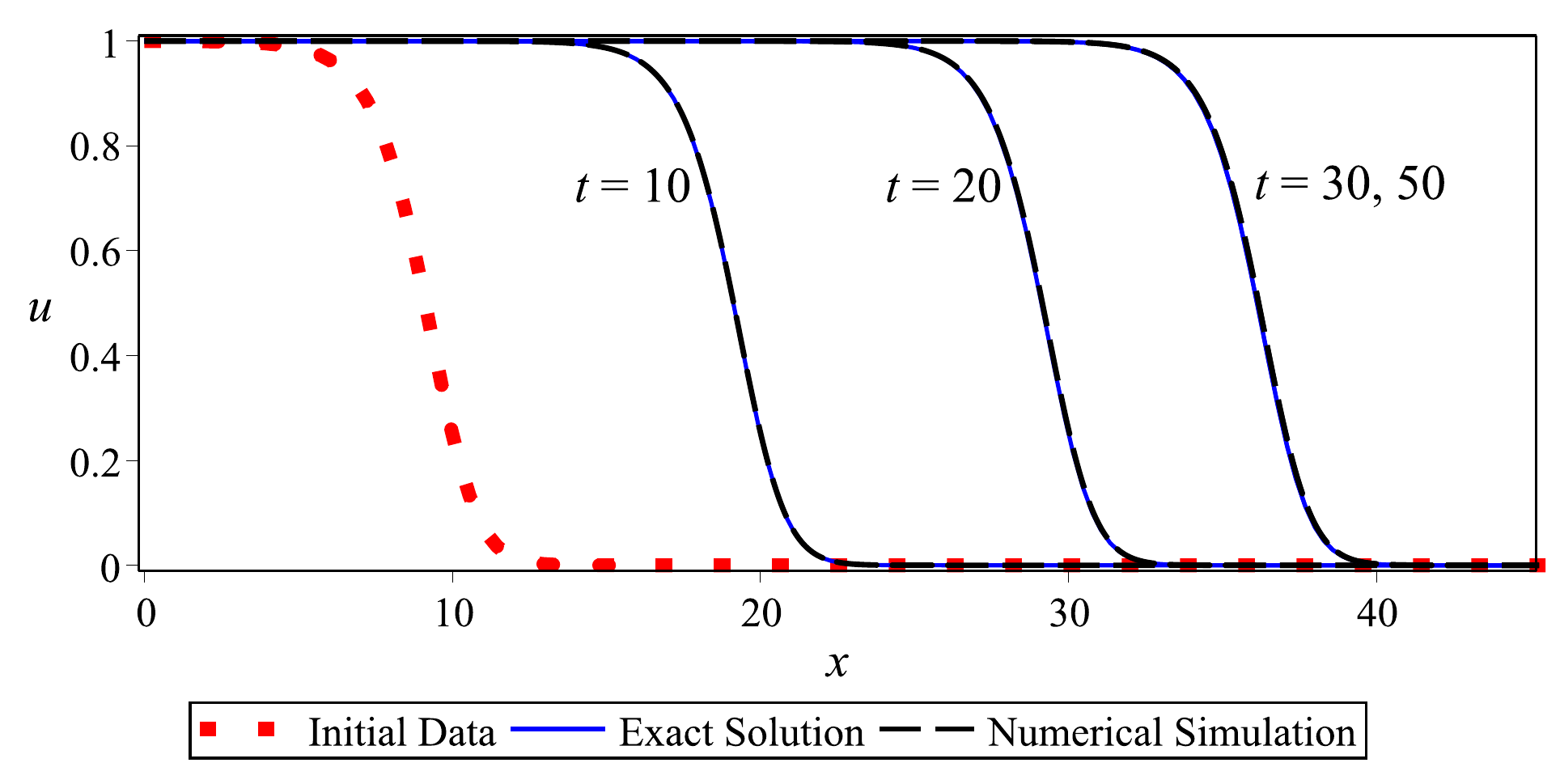}
\vspace{-0.2in}
\caption{Managed Fisher-KPP wavefronts for cases where the wave is slowed, stopped, and then sent on its way (top) or stopped and held stationary for all time (bottom). The left panels show the exact solution \eqref{fishersoln} plotted over space and time. The right panels show both the exact solution \eqref{fishersoln} and the numerical simulation at indicated times. In the top row, the wavefront is stopped for a while, before being sent along at the original wavespeed, with $C(t) = \frac{1}{2}\int_0^t  [2 - \tanh\left(\xi-17\right) + \tanh\left(\xi-33\right)] \mathrm{d}\xi$ and management parameters \eqref{fisherparams}. In the bottom row, the wavefront is stopped and frozen in place for all remaining time, using the management parameters \eqref{fisherparams} with $C(t) = \frac{1}{2}\int_0^t [1-\tanh(\xi-27)]\mathrm{d}\xi$. Colors in left panels are as indicated in Fig. \ref{Fig1}. \label{FigNumFisher}}
\end{center}
\end{figure}

In addition to nonlinear wave equations, travelling waves are commonly studied in the context of reaction-diffusion systems. Unlike solitary waves having one or more discrete peaks, the most commonly encountered traveling waves in the reaction-diffusion setting are moving sigmoidal wavefronts which comprise heteroclinic connections between two constant states \cite{grindrod1996theory}. One such reaction-diffusion equation is the Fisher-KPP equation \cite{fisher1937wave}, and a managed form of this equation reads
\begin{equation}\label{fisher}
\frac{\partial u}{\partial t} - \delta(t) \frac{\partial^2 u}{\partial x^2} - \gamma(t) u (1-u) =0\,. 
\end{equation}
The standard Fisher-KPP equation admits traveling wavefronts scaling like $u \sim \left( 1 + \exp(x-ct+x_0)\right)^{-2}$, as described in \cite{ablowitz1979explicit}. Assuming a similar type of solution for the non-autonomous equation \eqref{fisher} and placing this into \eqref{fisher}, we find that the constraint system \eqref{system} takes the form $4\delta(t) + \gamma(t) - 2\dot{C} =0$, $\gamma(t) - \delta(t) - \dot{C} =0$. Therefore, choosing the management parameters 
\begin{equation}\label{fisherparams}
\delta(t) = \frac{1}{5}\dot{C}(t)\,, \quad \gamma(t) = \frac{6}{5}\dot{C}(t)\,,
\end{equation}
the managed Fisher-KPP equation \eqref{fisher} has the solution
\begin{equation}\label{fishersoln}
u(x,t) = \left(1+ \exp(x-C(t)+x_0)\right)^{-2}.
\end{equation}
We numerically simulate \cite{numerics} the managed Fisher-KPP equation \eqref{fisher}, and compare the resulting solutions with the exact solution \eqref{fishersoln}, in Fig. \ref{FigNumFisher}. We first choose a management technique which takes a traveling wavefront, stops it for a while, and then sends it along at the original wavespeed, and a second management technique which takes a traveling wavefront with wavespeed of $1$ and gradually stops it after $t\approx 27$. Although \eqref{fisher} is dissipative rather than conservative like the KdV equation \eqref{kdvmanaged}, the numerical simulations show excellent agreement with the corresponding exact solutions \eqref{fishersoln}, supporting our theoretical claim that waves are robust and stable under controlled motion via soliton management.

In addition to traveling wavefronts emerging from scalar reaction-diffusion systems, it is possible to apply wave management to systems of equations. To illustrate this, consider a managed form of the Lotka-Volterra competition model \cite{okubo1989spatial}
\begin{equation} \label{fs}
\begin{aligned}
\frac{\partial  u_1}{\partial  t} & - \delta_1(t) \frac{\partial ^2 u_1}{\partial  x^2} - \gamma_1(t) u_1 (1-u_1 -\alpha u_2) =0\,,\\
\frac{\partial  u_2}{\partial  t} & - \delta_2(t) \frac{\partial ^2 u_2}{\partial  x^2} - \gamma_2(t) u_1 (1-\beta u_1 - u_2) =0\,,
\end{aligned}
\end{equation}
which arises in the competition between different populations for the same resources or predator-prey interactions. Here $\alpha$ and $\beta$ are constants that are not equal to one. The autonomous form of \eqref{fs} is known to admit solutions of the form $u_1 = \left( 1 + \exp(x-c_0 t+x_0)\right)^{-2}$ and $u_2 = 1-u_1$ (see \cite{krause2020non, kudryashov2015analytical, morita2009entire}), and this solution models the invasion of the habitat of one species by the other species (either $u_1$ or $u_2$ may be the invading species, depending upon the sign of the wavespeed $c_0$). Making a similar assumption on the form of a solution pair for the non-autonomous managed case by replacing $c_0t$ with $C(t)$, equation \eqref{fs} has a solution of the desired form when the constraint system \eqref{system} takes the form
\begin{subequations}\label{LVsystem}\begin{align}
\delta_1(t) - (1-\alpha) \gamma_1(t) + \frac{\mathrm{d}C}{\mathrm{d}t} &=0,\\
 (1-\alpha)\gamma_1(t) + 4\delta_1(t) - 2\frac{\mathrm{d}C}{\mathrm{d}t} &=0, \\
 \delta_2(t) + (1-\beta) \gamma_2(t) + \frac{\mathrm{d}C}{\mathrm{d}t} &=0,\\
 (1-\beta)\gamma_2(t) - 4\delta_2(t) + 2\frac{\mathrm{d}C}{\mathrm{d}t} &=0.
\end{align}\end{subequations}
Solving the constraint system \eqref{LVsystem}, the management parameters are found to be 
\begin{equation} \begin{aligned}
\delta_1(t)=\delta_2(t) = \frac{1}{5}\frac{\mathrm{d}C}{\mathrm{d}t}\,, ~~
\gamma_1(t) = \frac{6}{5(1-\alpha)}\frac{\mathrm{d}C}{\mathrm{d}t}\,,\\
 \gamma_2(t) = -\frac{6}{5(1-\beta)}\frac{\mathrm{d}C}{\mathrm{d}t}\,.
\end{aligned}\end{equation}
The managed Lotka-Volterra competition model \eqref{fs} then has the exact solution
\begin{equation}\begin{aligned}
u_1(x,t) & = \frac{1}{\left(1+ \exp(x-C(t)+x_0)\right)^{2}}\,,\\
u_2(x,t) & = 1 - \frac{1}{\left(1+ \exp(x-C(t)+x_0)\right)^{2}}\,.
\end{aligned}\end{equation} 
This kind of solution may be useful to model ecological interventions aimed at allowing populations to migrate through an ecosystem under climate change. 

\section{Control of the Benjamin-Bona-Mahony equation}
All of the examples considered thus far result in constraint systems \eqref{system} which yield management parameters which are proportional to $\dot{C}$, and one might think this is always the case. However, the situation changes if one or more of the operators $L$ or $N$ involve time derivatives. To illustrate this point, we consider the Benjamin-Bona-Mahony (BBM) equation \cite{benjamin1972model}. 

The managed form of the Benjamin-Bona-Mahony equation reads
\be \label{BBM}
\frac{\pa u}{\pa t}  - \delta(t) \frac{\pa^3 u}{\pa t \pa x^2} + \gamma(t)(1+u)\frac{\pa u}{\pa x}=0\,.
\ee
Motivated by exact solutions of \cite{olver1979euler} for the autonomous form of the BBM equation, we assume $U(Z)=\text{sech}^2\left( Z\right)$, finding that the system \eqref{system} gives 
\be 
12\frac{\mathrm{d}C}{\mathrm{d}t}\delta(t) - \gamma(t) =0, \quad (4\delta(t) -1)\frac{\mathrm{d}C}{\mathrm{d}t} + \gamma(t)=0.
\ee
Solving this system we obtain the management parameters
\be \label{BBMcontrol}
\delta(t) = \frac{1}{16}\,, \quad \gamma(t) = \frac{3}{4}\frac{\mathrm{d}C}{\mathrm{d}t}\,.
\ee
Therefore, \eqref{BBM} has the soliton solution 
\be 
u(x,t) = \text{sech}^2\left( x-C(t)-x_0\right),
\ee
provided that the control parameters satisfy \eqref{BBMcontrol}. Unlike other cases, note that the soliton management term is constant, and only the nonlinearity management term evolves in time. This difference is due to the fact that the BBM equation involves a time derivative in the dispersion term. Despite this fundamental difference in the structure of the BBM equation, our control technique still applies.

\section{Linear versus nonlinear wave equations}\label{seclinearwave}
Although the wave management technique is designed for the control of solitary waves or wavefronts in nonlinear equations, it is natural to wonder whether it can be used to control linear waves. To this end, consider the managed linear wave equation 
\begin{equation}   \label{linear0}
\frac{\partial^2 u}{\partial t^2} = \delta(t) \dfrac{\partial^2 u}{\partial x^2}\,,
\end{equation}  
subject to the initial data
\begin{equation}   \label{linearwave2}
u = f(x) \quad \text{and} \quad \frac{\partial u}{\partial t} = g(x) \quad \text{at} \quad t=0\,.
\end{equation}  
Although the problem \eqref{linear0} may look superficially distinct from a PDE of the form \eqref{managedpde}, note that we may always write \eqref{linear0} as the system 
\be 
\dfrac{\partial u}{\partial t} = v \quad \text{and} \quad \dfrac{\partial v}{\partial t} = \delta(t) \dfrac{\partial^2 u}{\partial x^2}\,.
\ee
In particular, this means that scalar equations with more than one time derivative still fall under the control framework of Section \ref{secgeneral}.

The standard linear wave equation on $\mathbb{R}$ admits waves of the form $U(x\pm ct)$ for generic envelope $U$. The functional form of $U$ is then determined by the initial data. If we assume a solution $u = U(x-C(t))=U(Z)$, we find
\begin{equation}   \label{mgmtfail}
\left\lbrace \left( \dfrac{\mathrm{d}C}{\mathrm{d}t}\right)^2 - \delta(t)\right\rbrace \dfrac{\mathrm{d}^2 U}{\mathrm{d}Z^2} -  \dfrac{\mathrm{d}^2 C}{\mathrm{d}t^2}   \dfrac{\mathrm{d} U}{\mathrm{d}Z} =0\,.
\end{equation}  
Even if we choose $\delta(t) =  \left( \frac{\mathrm{d}C}{\mathrm{d}t}\right)^2$, the second term persists and hence the only possibility is either $U = \text{constant}$ or $C(t)$ is a linear function (which is just the standard case of a constant wavespeed $c$, $C(t) = ct$). As such, the standard linear wave equation is not controllable under wave management.

From \eqref{mgmtfail} it is clear that \eqref{linear0} does not have a sufficient number of degrees of freedom to allow us to manage the wavespeed: Although it has a dispersion term, it lacks a loss/gain type term. Consider, then, the dissipative wave equation
\begin{equation}   \label{linearwave1}
\frac{\partial^2 u}{\partial t^2} + \gamma(t) \frac{\partial u}{\partial t} = \delta(t) \dfrac{\partial^2 u}{\partial x^2}\,,
\end{equation}  
subject again to the initial data \eqref{linearwave2}. Here $\gamma(t)$ plays the role of a loss or gain parameter, resulting in a de-amplification ($\gamma(t)>0$) or amplification ($\gamma(t) <0$) of a travelling wave in the autonomous setting. Assuming solutions of the form $u(x,t) = U(x\pm C(t))$, the constraint system \eqref{system} takes the from 
\begin{equation}   \label{linearwave3}
\left(\dfrac{\mathrm{d}C}{\mathrm{d}t}\right)^2 - \delta(t)=0\,, \quad \dfrac{\mathrm{d}^2 C}{\mathrm{d}t^2} + \gamma(t)\dfrac{\mathrm{d}C}{\mathrm{d}t}=0 \,.
\end{equation}  
Choosing management parameters
\begin{equation}   \label{linearwave4}
\delta(t) = \left(\dfrac{\mathrm{d}C}{\mathrm{d}t}\right)^2\,,\quad \gamma(t) = - \dfrac{\mathrm{d}}{\mathrm{d}t} \log \left| \dfrac{\mathrm{d}C}{\mathrm{d}t}\right|\,,
\end{equation}  
we have that any $u(x,t) = U(x\pm C(t))$ is a solution of the PDE \eqref{linearwave1}. The solution of \eqref{linearwave1} then takes the form $u(x,t) = U_-(x-C(t))+U_+(x+C(t))$. Taking into account the initial data \eqref{linearwave2}, from d'Alembert's formula we have the exact solution to \eqref{linearwave1}:
\begin{equation}\begin{aligned}   \label{frictionwaves1}
u(x,t) & = \dfrac{f(x-C(t))+f(x+C(t))}{2} \\
& \qquad + \dfrac{1}{2}\left(\dfrac{\mathrm{d}C}{\mathrm{d}t}(0)\right)^{-1}\int_{x-C(t)}^{x+C(t)} g(\xi)\mathrm{d}\xi \,.
\end{aligned}\end{equation}  

Although the wave equation \eqref{linearwave1} is linear, the wave management works since there is both a dispersion and a loss/gain term. The fact that both of these terms are linear does not preclude application of wave management. Note that the parameter $\gamma(t)$ is only present when the wavespeed is changing. On time intervals when the wavespeed is constant, we have from \eqref{linearwave4} that $\gamma(t)\equiv 0$. Therefore, this dissipative control term is only present when one is actively adjusting the motion of a wave, with the term vanishing when the wave is no longer being controlled. 

\section{Nonlinear Klein-Gordon equations}
Motivated by the dissipative linear wave equation in Section \ref{seclinearwave}, we will now show how to control the motion of a \textit{nonlinear} form of this equation, namely a nonlinear Klein-Gordon equation \cite{scott1969nonlinear, morawetz1968time, grundland1992family}. Dissipative nonlinear Klein-Gordon equations have recently been linked to emergent spatial structure, including waves and patterns \cite{zemskov2016diffusive,ritchie2022turing}. To this end, we consider a dissipative analogue of the Klein-Gordon equation with cubic nonlinearity, 
\begin{equation}  \label{KG1}
\frac{\partial^2 u}{\partial t^2}  = \delta(t) \dfrac{\partial^2 u}{\partial x^2} + \gamma(t) \left\lbrace u - 2 u^3 - \frac{\partial u}{\partial t} \right\rbrace\,.
\end{equation}  
Here we have included a time derivative in the operator $N[u]$. We will consider a solution of the form of the solitary wave $u(x,t)=U(Z)=\text{sech}(Z)$, with the system \eqref{system} then found to be
\be \label{KG2}
\left(\dfrac{\mathrm{d}C}{\mathrm{d}t}\right)^2 - \delta(t) - \gamma(t) =0\,, \quad \dfrac{\mathrm{d}^2 C}{\mathrm{d}t^2} + \gamma(t)\dfrac{\mathrm{d}C}{\mathrm{d}t}=0\,.
\ee
Note that the constraint system \eqref{KG2} is very similar to the constraint system for the linear dissipative wave equation \eqref{linearwave3}, with the difference being that the dispersion term $\delta(t)$ is now coupled to the loss/gain term $\gamma(t)$ due to the nonlinearity in \eqref{KG1}. Choosing management parameters
\begin{equation}   \label{KG3}
\delta(t) = \left(\dfrac{\mathrm{d}C}{\mathrm{d}t}\right)^2 + \dfrac{\mathrm{d}}{\mathrm{d}t} \log \left| \dfrac{\mathrm{d}C}{\mathrm{d}t}\right|\,,\quad \gamma(t) = - \dfrac{\mathrm{d}}{\mathrm{d}t} \log \left| \dfrac{\mathrm{d}C}{\mathrm{d}t}\right|\,,
\end{equation}  
we have that the solitary wave
\be 
u(x,t) =  \text{sech}\left( x-C(t)-x_0\right)
\ee
is a solution to the managed form of the dissipative Klein-Gordon equation \eqref{KG1}. As was true for the linear wave equation, here the control parameter $\gamma(t)$ is zero when the wave motion is not being actively controlled. We have grouped the nonlinear terms with the first time derivative, as well, and this means that the nonlinear terms play no role unless the wave motion is being changed. Of course, one could very easily change the problem to include two parameters, say $\gamma_1(t)$ multiplying the nonlinearity and $\gamma_2(t)$ controlling the first time derivative. In this case, we will obtain a constraint equation for $\gamma_2(t)$ (which will take the form of the second equation in \eqref{KG2}) and a separate equation linking $\delta(t)$ and $\gamma_1(t)$. The latter of these can be solved for one of $\delta(t)$ or $\gamma_1(t)$, with the other remaining a free parameter to be used for other means (for instance, for changing the shape of the solitary wave, if that is desirable).

\section{Discussion}
We have described a method for controlling the motion (speed and direction) of a nonlinear wave -- while keeping its shape or structure unchanged -- through a form of soliton management. While dispersion management and soliton management have seen a number of applications over the last three decades, standard applications often distort the wave envelope in some manner. Our approach to soliton management keeps the wave envelope unchanged, and hence preserves adiabaticity. In light of the analysis and examples provided, this form of soliton management appears to be a generic, useful, and relatively straightforward strategy for controlling the motion of a traveling wave as desired. 

Any method will have limitations, and soliton management is no different. 
To successfully apply wave management, one or both of the dispersion and loss/gain terms should be controllable in time, and such controllability will depend upon the specific physical system under consideration. This restriction is common to all implementations of soliton management, not only our approach. Although this potential lack of controllability will limit some experimental implementations, there are still many applications lending themselves to this manner of dispersion or loss/gain control, including examples in optics, atomic physics, condensed matter physics, and problems involving heat and mass transfer. Many of these systems are modeled using integrable or conditionally integrable (integrable only for certain parameter sets) systems. Indeed, all integrable or conditionally integrable systems have this restrictive property; they all require very precise parameters for a given nonlinear wave to exist. Therefore, we do not view the fact that the management parameters need to satisfy the constraint system \eqref{system} as particularly limiting relative to standard approaches for finding solitary waves in integrable or conditionally integrable systems. The specific examples included herein are but a small sample of what is possible, and we expect that the method will work well for many integrable systems.

Our application of soliton management is relevant to waves moving along one direction, with wave coordinate $\mathbf{k}\cdot \mathbf{x}-C(t)$. This is the geometric configuration that the vast majority of solitary waves and wavefronts satisfy. Waves that are fundamentally higher dimensional are also possible, appearing in a variety of applications, although exact solutions do not tend to exist due to a typical lack of integrability in more than one space dimensions. Still, in addition to numerical simulations, there are a variety of methods, such as the variational approximation, which are used to study these systems. It would be a worthwhile topic for future work to extend the philosophy of the present paper in order to develop tools for the manipulation and control of higher-dimensional waves which preserve the structure of said waves. In theory, one could use a variational approximation in place of an exact solution, and apply our method while ensuring that the envelope of the variational approximation was unchanged in time. Application of a time-varying potential to move a matter wave in a 2D or 3D atomic BEC was recently explored in \cite{van2021time}, where it was commented that one must be careful with this motion to preserve adiabaticity. A control method ensuring that such a matter wave can be moved, while preserving adiabaticity of the BEC ground state, will be considered in future work.


\end{document}